\documentclass[twocolumn,prd,aps,longbibliography,superscriptaddress,preprintnumbers,tightenlines,nofootinbib,
amsfonts,amsmath,showkeys]{revtex4-2}

\pdfoutput=1

\usepackage{graphicx}
\usepackage{amsmath,amssymb,mathrsfs}
\usepackage{amsfonts}
\usepackage[usenames,dvipsnames]{xcolor}
\usepackage{xcolor}
\usepackage{wasysym}
\usepackage[colorlinks]{hyperref}
\usepackage{tensor}
\usepackage{bm}
\usepackage[english]{babel}
\usepackage[utf8]{inputenc}
\usepackage{booktabs}

\graphicspath{{pics/}}

\newcommand{\AEI}{Max Planck Institute for Gravitational Physics (Albert Einstein Institute), Am M\"uhlenberg 1, Potsdam 14476, Germany}
\newcommand{\NBIA}{Niels Bohr International Academy, Niels Bohr Institute, Blegdamsvej 17, 2100 Copenhagen, Denmark}
\newcommand{\UCD}{School of Mathematics and Statistics, University College Dublin, Belfield, Dublin 4, D04 V1W8, Ireland}
\newcommand{\UTA}{Center of Gravitational Physics, University of Texas at Austin, Austin, Texas 78712, USA}
\newcommand{\UoS}{School of Mathematical Sciences and STAG Research Centre, University of Southampton, Southampton, SO17 1BJ, United Kingdom}
\newcommand{\Maryland}{Department of Physics, University of Maryland, College Park, MD 20742, USA}

\definecolor{dodgerblue}{HTML}{1E90FF}
\definecolor{viennared}{HTML}{DA0A14}

\hypersetup{citecolor=dodgerblue}

\newcommand{\abs}[1]{\left\lvert#1\right\rvert}

\renewcommand{\d}[2]{\frac{\operatorname{d}\!#1}{\operatorname{d}\!#2}}	

\newcommand{\mr}[0]{\epsilon} 
\newcommand{\lmr}[0]{q} 
\newcommand{\smr}[0]{\nu} 
\newcommand{\flux}{\mathcal{F}} 

\newcommand{\SEOBNRvVHM}{\texttt{SEOBNRv5HM}}

\newcommand{\SEOBNR}[1]{\texttt{SEOBNR#1}}

\begin{document}

\title{Enhancing the \SEOBNR{v5} effective-one-body waveform model with \\ second-order  gravitational self-force fluxes}

\author{Maarten van de Meent} \affiliation{\NBIA}\affiliation{\AEI}
\author{Alessandra Buonanno} \affiliation{\AEI}\affiliation{\Maryland}
\author{Deyan P. Mihaylov} \affiliation{\AEI}
\author{Serguei Ossokine} \affiliation{\AEI}
\author{Lorenzo Pompili} \affiliation{\AEI}
\author{Niels Warburton} \affiliation{\UCD}
\author{Adam Pound} \affiliation{\UoS}
\author{Barry Wardell} \affiliation{\UCD}
\author{Leanne Durkan} \affiliation{\UCD}\affiliation{\UTA}
\author{Jeremy Miller} \affiliation{Department of Physics, Ariel University, Ariel 40700, Israel}

\date{\today}

\begin{abstract}
  We leverage recent breakthrough calculations using second-order
  gravitational self-force (2GSF) theory to improve both the
  gravitational-mode amplitudes and radiation-reaction force in
  effective-one-body~(EOB) waveform models. We achieve this by introducing
  new calibration parameters in the {\tt SEOBNRv5HM} mode amplitudes, and
  matching them to the newly available 2GSF energy-flux multipolar data for
  quasicircular nonspinning binary black holes. We find that this significantly
  improves the {\tt SEOBNRv5HM} energy flux, when compared to numerical-relativity
  (NR) simulations of binary black holes with mass ratios between 1:1 and 1:20. Moreover, we find that,
  once the conservative part of the {\tt SEOBNRv5} dynamics is calibrated,
  the {\tt SEOBNRv5HM} waveform model with 2GSF information  reproduces the
  binding energy of NR simulations more accurately, providing a
  powerful check of the consistency and naturalness of the EOB approach. While we only
  include nonspinning 2GSF information, the more accurate binding
  energy and energy flux carry over to the {\tt SEOBNRv5} waveform models for spinning binary black holes. Thus,
our results improve the latest generation of {\tt SEOBNR} waveform models (i.e., {\tt SEOBNRv5}),
which has been recently completed for use in the upcoming fourth observing (O4) run of the LIGO-Virgo-KAGRA Collaboration.
\end{abstract}

\maketitle

\section{Introduction} \label{sec:Intro}

During their first, second and third observing runs~\cite{TheLIGOScientific:2016pea, LIGOScientific:2018mvr, LIGOScientific:2020ibl,LIGOScientific:2021usb,LIGOScientific:2021djp}, the LIGO \cite{TheLIGOScientific:2014jea} and Virgo \cite{TheVirgo:2014hva} gravitational-wave (GW) observatories have detected GWs from about ninety coalescences of compact binaries, composed of black holes (BHs)
and/or neutron stars. Moreover, independent confirmations of these detections, as well as claims of new
ones, were obtained in Refs.~\cite{Nitz:2018imz,Nitz:2019hdf,Venumadhav:2019lyq,Zackay:2019btq,Nitz:2021zwj,Olsen:2022pin}. All together, these results have firmly established the field of GW astronomy. The vast majority of the observed GW signals involve binaries with comparable masses~\cite{LIGOScientific:2018jsj, Venumadhav:2019lyq, LIGOScientific:2020ibl,Nitz:2021zwj} although a few signals show evidence for binaries with greater mass asymmetry \cite{LIGOScientific:2020stg,LIGOScientific:2020zkf}, quantified by the ratio of the component masses ($\lmr=m_1/m_2 \geq 1$ or $\mr=m_2/m_1 \leq 1$, $m_1$ and $m_2$ being the primary and secondary masses in the binary, respectively).

As the number of GW detections is expected to increase in upcoming observation runs \cite{KAGRA:2013rdx,LIGOScientific:2018jsj,LIGOScientific:2020kqk}, so likely will the number of  asymmetric coalescences. It is therefore important that waveform models used for detecting, identifying, and analyzing the GW signals faithfully represent binaries in the small-mass-ratio (SMR) regime.

Effective-one-body (EOB)
theory~\cite{Buonanno:1998gg,Buonanno:2000ef,Damour:2000we,Damour:2001tu,Buonanno:2005xu}
provides waveform models that can be used for the analysis of GW
signals by combining results from various first-principle methods for
solving the two-body problem in General Relativity, such as
post-Newtonian (PN) theory~\cite{Blanchet:2013haa,Schafer:2018kuf,Porto:2016pyg,Levi:2018nxp}
and numerical relativity (NR)~\cite{Pretorius:2005gq,Campanelli:2005dd,Baker:2005vv}. Moreover,
EOB waveform models are constructed in such a way that they reduce to
test-body motion around a black hole in the limit of vanishing
$\epsilon$. Thus, to improve fidelity of EOB waveform models, it
is in principle straightforward and natural to incorporate results from SMR perturbation theory
or gravitational self-force (GSF) theory~\cite{Poisson:2011nh} in the EOB formalism~\cite{Damour:2009sm}.

There are two main families of EOB models, {\tt SEOBNR} (e.g., see Refs.~\cite{Bohe:2016gbl,Cotesta:2018fcv,Ossokine:2020kjp}) and {\tt TEOBResumS} (e.g., see Refs.~\cite{Nagar:2018zoe,Nagar:2020pcj,Gamba:2021ydi}). We consider here the former, and in particular we focus on the latest generation of {\tt SEOBNR} models (i.e., {\tt SEOBNRv5}%
\footnote{The \texttt{SEOBNRv5} family of models is publicly available through the Python package \texttt{pySEOBNR}:\\ \href{https://git.ligo.org/waveforms/software/pyseobnr}{https://git.ligo.org/waveforms/software/pyseobnr}.
Stable versions of \texttt{pySEOBNR} are published through the Python Package Index (PyPI), and can be installed via ~\texttt{pip install pyseobnr}.}%
) recently developed  in Refs.~\cite{Khalilv5,Pompiliv5,Ramos-Buadesv5,Mihaylovv5}  for the upcoming
fourth observing run of the LIGO-Virgo-KAGRA (LVK) Collaboration, which will also include the KAGRA~\cite{KAGRA:2018plz,KAGRA:2020tym} detector in Japan.

The inclusion of GSF results in EOB waveforms has, so far, been limited to the inclusion of higher PN test-body
coefficients in the energy flux and gravitational-mode amplitudes (e.g., see
Refs.~\cite{Bohe:2016gbl,Cotesta:2018fcv,Nagar:2018zoe,Nagar:2020pcj}), fits of EOB-mode amplitudes to the inspiral Teukolsky-multipolar modes~\cite{Yunes:2009ef,Yunes:2010zj} and the calibration of EOB Hamiltonians~\cite{Taracchini:2013rva,Bohe:2016gbl} to match the first-order GSF correction to the nonspinning shift of the innermost stable circular orbit (ISCO)~\cite{Barack:2009ey}.
Various studies~\cite{Barausse:2011dq,LeTiec:2011dp,Akcay:2012ea,Antonelli:2019fmq,Akcay:2015pjz}
considered improving the EOB Hamiltonian by incorporating GSF
corrections to the binding energy through use of the first law of
binary mechanics~\cite{LeTiec:2011ab,LeTiec:2015kgg}. In the standard
gauge~\cite{Buonanno:1998gg,Damour:2000we} used in EOB waveform models, this leads to a gauge singularity at the
lightring (or photon orbit) radius~\cite{Akcay:2012ea,Antonelli:2019fmq}. This
singularity can be cured by reformulating the EOB Hamiltonian in a
different gauge~\cite{Antonelli:2019fmq,Damour:2017zjx}. So far, this
has not been implemented in any fully featured EOB waveform model employed for
LVK analyses. Recently, a version of the \texttt{TEOBResumS} model has been
produced~\cite{Nagar:2022fep} that incorporates most of the previously
calculated GSF corrections to the EOB Hamiltonian without, however, calibrating it
to NR simulations, and focusing only on the inspiral phase to avoid issues with the
lightring divergence.

Recent calculations in GSF theory have provided the second-order GSF (henceforth, 2GSF) correction to the
energy flux~\cite{Warburton:2021kwk} as well as corresponding post-adiabatic
waveforms~\cite{Wardell:2021fyy}. References~\cite{Albertini:2022rfe,Albertini:2022dmc}
carried out a detailed comparison of these results against NR waveforms using a version of the {\tt TEOBResumS}
waveforms; among other things, 2GSF waveforms enabled a
precise assessment of the accuracy of the {\tt TEOBResumS} family
in the SMR regime. However, these references did not seek to
incorporate the 2GSF data into EOB waveform models. In this
work we will capitalize on the 2GSF breakthrough by directly
incorporating 2GSF energy flux corrections in the latest generation of {\tt SEOBNR}
models. We will see that, quite interestingly, including these corrections
does not only improve the waveform models at small mass-ratios, but also for
comparable masses.

In this paper, we employ units such that $G=c=1$. The component masses of a binary are denoted $m_1$ and $m_2$ with $m_1\geq m_2$. The total mass is $M=m_1+m_2$, the reduced mass is $\mu= m_1m_2/M$, while $\smr$ denotes the symmetric mass-ratio $\mu/M$.

\section{Basics of the effective-one-body approach}\label{sec:EOB}

In the EOB formalism the dynamics of a compact binary is mapped onto that of an effective test mass (or test spin) in a deformed BH background,
with the deformation parameter being the symmetric mass-ratio.
The EOB approach builds semi-analytical inspiral-merger-ringdown waveforms by combining analytical predictions for the inspiral
and ringdown phases (from BH perturbation theory) with physically-motivated ansatzes
for the plunge-merger stage. The EOB  waveforms are then made highly accurate via a calibration to NR waveforms. The EOB
formalism relies on three key ingredients: the EOB conservative dynamics (i.e., a two-body Hamiltonian), the EOB radiation-reaction forces
(i.e., the energy flux) and the EOB gravitational modes. In this paper we shall limit to the inspiral-portion of the coalescence of nonspinning BHs.
Here, we describe each of the main EOB ingredients, as necessary (e.g., see Ref.~\cite{Pompiliv5} for details).

\subsection{EOB Hamiltonian}

In the binary's center-of-mass frame, the motion is described by the orbital phase $\phi$, the relative position $r$, the radial momentum
$p_r$ and the angular momentum $p_\phi$. In the EOB formalism, the Hamiltonian $H_\text{EOB}$, describing the conservative binary dynamics,
is related to the effective Hamiltonian $H_\text{eff}$, describing the dynamics of a test body in a deformed BH background, via the energy map~\cite{Buonanno:1998gg}
\begin{equation}
\label{EOBmap}
H_\text{EOB} = M \sqrt{1 + 2 \nu \left(\frac{H_\text{eff}}{\mu} - 1\right)}\,.
\end{equation}
For nonspinning binaries, in the $\nu\to0$ limit, $H_\text{eff}$ reduces to the Hamiltonian of a (nonspinning) test mass in a Schwarzschild background.
The nonspinning EOB Hamiltonian was first derived in Refs.~\cite{Buonanno:1998gg,Buonanno:2000ef} with 2PN information. It was then extended to 3PN order in Ref.~\cite{Damour:2000we} and to 4PN order in Ref.~\cite{Damour:2015isa}. As of today, the 5PN~\cite{Bini:2019nra,Bini:2020wpo,Blumlein:2021txe} and 6PN terms~\cite{Bini:2020nsb,Bini:2020hmy} are partially known. In the non-spinning limit $H_{\rm eff}$ reads:
\begin{align}\label{eq:Heff}
 	H_{\rm eff} = \sqrt{p_{r_{*}}^2 + A(r) \Big[\mu^2 + \frac{p_\phi^2}{r^2} + Q(r,p_{r_{*}})\Big]},
\end{align}
where $p_{r_{*}}$ is the canonical momentum conjugate to the tortoise coordinate $r_{*}$,
\begin{align}
\d{r}{r_{*}} = \frac{p_{r_{*}}}{p_r} = A(r)\sqrt{\bar{D}(r)}.
\end{align}
The 5PN Taylor-expanded potential $A$ is given by
\begin{align}
\label{ApmTay}
A(u) &= 1 - 2 u + 2\nu u^3 + \nu \left(\frac{94}{3}-\frac{41 \pi ^2}{32}\right) u^4 \nonumber\\
&\quad
+ \bigg[\nu  \left(\frac{2275 \pi ^2}{512}-\frac{4237}{60}+\frac{128 \gamma_E }{5}+\frac{256 \ln 2}{5}\right) \nonumber\\
&\qquad
+ \left(\frac{41 \pi ^2}{32}-\frac{221}{6}\right) \nu ^2 +\frac{64}{5} \nu \ln u\bigg] u^5\nonumber\\
&\quad
 + \left[\nu a_6 - \nu\left(\frac{144 \nu}{5}+\frac{7004}{105}\right) \ln u\right] u^6,
\end{align}
where $u\equiv M/r$ and $\gamma_E\simeq 0.5772$ is the Euler gamma constant. In Eq.~(\ref{ApmTay}), except for the log part, we
replace the (partially known) coefficient of $u^6$ by the parameter $a_6$. The latter is used
in the construction of the {\tt SEOBNRv5} waveform models~\cite{Pompiliv5} (and also in previous
EOB families) to calibrate against NR simulations. Furthermore, the resummed form of $A$, the potentials
$\bar{D}$ and $Q$, and the spinning EOB Hamiltonian used for the calibration of the
{\tt SEOBNRv5} waveform model can be found in Refs.~\cite{Khalilv5,Pompiliv5}.

\subsection{EOB gravitational modes}\label{sec:EOB waveform}

The observer-frame's gravitational polarizations read
\begin{align}
		h(t;\iota,\varphi_0)&= h_{+}(t;\iota,\varphi_0) - i  h_{\times}(t;\iota,\varphi_0) \\
		\label{eq:waveform_harmonics}
	&= \sum_{\ell=2}^\infty \sum_{m=-\ell}^{\ell} {}_{-2}{Y}{_{\ell m}}(\iota,\varphi_0)\,h_{\ell m}(t),
\end{align}
where we denote with $\iota$ the binary's inclination angle (computed with respect to the
direction perpendicular to the orbital plane), $\varphi_0$ the azimuthal direction to the observer,
and $\tensor[_{-2}]{Y}{_{\ell m}}(\iota,\varphi_0)$'s the -2 spin-weighted spherical harmonics.
For nonspinning binaries $h_{\ell m}=(-1)^{\ell} h_{\ell-m}^*$, therefore one can restrict the discussion to $(\ell,m)$ modes with $m > 0$.

In the {\tt SEOBNRv5HM} waveform model, the 7 most dominant modes are included, specifically
$(\ell,m) = (2,2), (2,1), (3,3), (3,2), (4,4), (4,3),$ and $(5,5)$. The EOB modes for the entire coalescence (i.e., inspiral, plunge, merger and
ringdown (RD)), can be written as:
\begin{equation}
	h_{\ell m}(t)=
\begin{cases}
h_{\ell m}^{\text {insp-plunge }}(t), & t<t_{\text {match }}^{\ell m} \\ h_{\ell m}^{\text {merger-RD }}(t), & t>t_{\text {match }}^ {\ell m }
\end{cases},
\label{EOB:IMR}
\end{equation}
where $t_{\text {match }}^{\ell m}$ is defined as
\begin{equation}
	\label{eq:t_match}
	t_{\text {match }}^{\ell m}=
	\begin{cases}
	t_{\text {peak }}^{22}, &(\ell, m) = (2,2),(3,3),(2,1), \\
	& \qquad\quad\,\; (4,4),(3,2),(4,3) \\
	t_{\text {peak }}^{22}-10 M,  &(\ell, m) =(5,5),\end{cases}
\end{equation}
where $t_{\text {peak }}^{22}$ is the peak of the $(2,2)$-mode amplitude. The choice of a different attachment point for the $(5,5)$ mode is discussed in
Refs.~\cite{Cotesta:2018fcv,Pompiliv5}. In the \texttt{SEOBNRv5HM} model, one imposes that~\cite{Pompiliv5}
\begin{equation}
	\label{eq:t_attach}
	t_{\text{peak}}^{22} = t_{\rm{ISCO}}+ \Delta t^{22}_{\rm{ISCO}}\,,
\end{equation}
where $t_{\rm{ISCO}}$ is the time at which $r = r_{\rm{ISCO}}$, with $r_{\rm{ISCO}}$ the ISCO radius of a Kerr BH \cite{Bardeen:1972fi} with the final mass and spin of the
remnant object~\cite{Jimenez-Forteza:2016oae, Hofmann:2016yih}. The parameter
$\Delta t^{22}_{\rm{ISCO}}$ in Eq.~(\ref{eq:t_attach}) is the second EOB calibration parameter (the first being $a_6$
in the potential $A$ of Eq.~(\ref{ApmTay})), which is determined by minimizing the disagreement between EOB and NR
waveforms throughout the inspiral-plunge stage. Henceforth, we focus only on the $h_{\ell m}^{\rm insp-plunge}(t)$, and do not provide details on
how the $h_{\ell m}^{\rm merger-RD}(t)$ is constructed.

The inspiral-plunge EOB modes are written as
\begin{equation}
\label{insppl}
h_{\ell m}^{\text {insp-plunge}}(t)= h_{\ell m}^{\rm F}(t)\,N_{\ell m}(t)\,,
\end{equation}
where $h_{\ell m}^{\rm F}(t)$'s resum the PN-expanded GW modes for circular orbits in factorized form~\cite{Damour:2007xr,Damour:2007yf,Damour:2008gu,Pan:2010hz}, while $N_{\ell m}(t)$'s are the nonquasicircular (NQC) corrections~\cite{Damour:2002vi,Damour:2008te,Buonanno:2009qa}, aimed at incorporating relevant radial effects during the plunge (see below).
The factorized inspiral modes are given by~\cite{Damour:2008gu}
\begin{equation}\label{eq:hfactors}
	h_{\ell m}^F =
		h_{\ell m}^{\rm N}
		\hat{S}_{\ell m}
		T_{\ell m}
		f_{\ell m} e^{i \delta_{\ell m}}.
\end{equation}
The first factor, $h_{\ell m}^{\rm N}$, encodes the leading (Newtonian) order waveform, and its explicit expression is
\begin{equation}\label{eq:hnewt}
	h_{\ell m}^{\rm N} =\frac{\smr M}{d_L} n_{\ell m} c_{\ell+\mr_{\ell m}}(\smr)v_\phi^{\ell+\mr_{\ell m}}Y_{\ell-\mr_{\ell m},-m}\left(\frac{\pi}{2},\phi\right),
\end{equation}
where $d_L$ is the luminosity distance, $Y_{\ell m}$ is a scalar spherical harmonic, $\mr_{\ell m}$ is the parity of the mode,
\begin{equation}
	\mr_{\ell m} = \left\{\begin{aligned}
		&0, &&\text{$\ell+m$ is even},\\
		&1, &&\text{$\ell+m$ is odd},
	\end{aligned} \right.
\end{equation}
and the functions $n_{\ell m}$ and $c_{k}(\smr)$ are given by
\begin{equation}
	n_{\ell m} = \left\{\begin{aligned}
		& \frac{8\pi (i m)^\ell}{(2\ell+1)!!}\sqrt{\tfrac{(\ell+1)(\ell+2)}{\ell(\ell-1)}}, &&\text{$\ell+m$ is even},\\
		& \frac{-16i \pi (i m)^\ell}{(2\ell+1)!!}\sqrt{\tfrac{(2\ell+1)(\ell+2)(\ell^2-m^2)}{(2\ell-1)(\ell+1)\ell(\ell-1)}}, &&\text{$\ell+m$ is odd},
	\end{aligned} \right.
\end{equation}
and
\begin{equation}
	c_{k}(\smr) = \left(\tfrac{1-\sqrt{1-4\smr}}{2} \right)^{k-1}+(-1)^k \left(\tfrac{1+\sqrt{1-4\smr}}{2} \right)^{k-1}.
\end{equation}
Finally, $v_\phi$ in Eq.~\eqref{eq:hnewt} is given by
\begin{equation}
\label{vp}
	v_\phi = M\Omega r_\Omega,
\end{equation}
with
\begin{align}
\Omega &=\frac{\partial H_{\rm EOB}}{\partial p_\phi}, & \quad \frac{1}{r_\Omega^{3/2}} &= \left.\frac{\partial H_{\rm EOB}}{\partial p_\phi}\right|_{{p_r=0}}.
\end{align}
The second factor in Eq.~\eqref{eq:hfactors}, $\hat{S}_{\ell m}$, is an effective source term. Depending on the parity of the mode, it is given by
\begin{equation}\label{eq:Slm}
	\hat{S}_{\ell m} = \left\{\begin{aligned}
		&H_{\rm eff}, &&\text{$\ell+m$ is even},\\
		&L_{\rm eff} = v_\Omega p_\phi, &&\text{$\ell+m$ is odd},
	\end{aligned} \right.
\end{equation}
with
\begin{equation}
\label{vO}
	v_\Omega = (M\Omega)^{1/3}.
\end{equation}

The third factor in Eq.~\eqref{eq:hfactors}, $T_{\ell m}$, is an analytic resummation of the leading-order tail terms in the PN expansion~\cite{Blanchet:1997jj},
\begin{equation}\label{eq:Tlm}
	T_{\ell m} = \frac{\Gamma(\ell+1-2i\hat{\Omega})}{\Gamma(\ell+1)}e^{\pi m \hat{\Omega}} \left(\frac{4m M\Omega}{\sqrt{e}}\right)^{2i m\hat{\Omega}},
\end{equation}
where $\hat{\Omega} = \Omega H_{\rm EOB}$ is the orbital frequency normalized by the total energy.

The final two factors in Eq.~\eqref{eq:hfactors} encode additional physical information included in the waveform from PN theory. The function $f_{lm}$ is expanded as
\begin{equation}\label{eq:flm}
	f_{lm} = \left\{\begin{aligned}
		&(\rho_{\ell m})^\ell, &&\text{$m$ is even}\\
		&(\rho_{\ell m})^\ell +	f_{lm}^{\rm S}, &&\text{$m$ is odd.}
	\end{aligned} \right.
\end{equation}
The functions $\delta_{\ell m}$, $\rho_{\ell m}$, and $f_{lm}^{\rm S}$ are fixed by requiring that the waveform at fixed orbital frequency $\Omega$ matches known analytical PN and test-body expressions. The full expressions used in {\tt SEOBNRv5HM} are given in Appendix~B of~\cite{Pompiliv5}. In this work, we will further use  $\rho_{\ell m}$ to incorporate the new 2GSF energy flux results.

\subsection{EOB equations of motion and radiation-reaction force}

The EOB equations of motion read
\begin{subequations}
	\begin{align}
	\d{r}{t} 		&= \frac{\partial H_{\rm EOB}}{\partial p_r}, &
	\d{p_r}{t}		&= -\frac{\partial H_{\rm EOB}}{\partial r}+{F}_r,\\
	\d{\phi}{t} 	&= \frac{\partial H_{\rm EOB}}{\partial p_\phi}, &
	\d{p_\phi}{t}	&= {F}_\phi,
\end{align}
\label{eq:EOB-EOMs}%
\end{subequations}
where the radiation-reaction (RR) forces ${F}_r$ and ${F}_\phi$ are given by~\cite{Buonanno:2005xu}
\begin{equation}
	{F}_r = -\frac{\flux^{\rm EOB}}{M\Omega}\,\frac{p_r}{p_\phi}\,,
	\quad \quad
	{F}_\phi = -\frac{\flux^{\rm EOB}}{M\Omega},
\end{equation}
where $\flux^{\rm EOB}$ is the energy flux, which is given as a sum over $(\ell,m)$ modes~\cite{Damour:2008gu},
\begin{equation}\label{eq:EOBflux}
	\flux^{\rm EOB} = \sum_{\ell=2}^{8} \sum_{m=1}^{\ell} \flux_{\ell m}^{\rm EOB}.
\end{equation}
Each of the $(\ell, m$)-mode contributions to the energy flux is obtained from the inspiral-plunge waveform modes,
assuming quasi-circular orbits
\begin{align}\label{eq:fluxmodes}
	\flux_{\ell m}^{\rm EOB} &= d_L^2\frac{(m M\Omega)^2}{8\pi} \abs{h^{\rm insp-plunge}_{\ell m}}^2.
\end{align}

As discussed in Sec.~\ref{sec:EOB waveform}, in the {\tt SEOBNRv5} model, as in other EOB variants, the inspiral-plunge quasi-circular waveform
is enhanced by the NQC corrections $N_{\ell m}$ (see Eq.~(\ref{insppl})), which for the {\tt SEOBNR} models take the form
\begin{align}
N_{\ell m} &= \left[1+ \frac{p_{r^{*}}^2}{(r\Omega)^2}
	\left( a_1^{h_{\ell m}} + \frac{a_2^{h_{\ell m}}}{r} + \frac{a_3^{h_{\ell m}}}{r^{3/2}} \right)
	\right]
	\nonumber \\
	&\qquad\times
\exp\left[i \left(b_1^{h_{\ell m}} \frac{p_{r^{*}}}{r\Omega} + b_2^{h_{\ell m}} \frac{p_{r^{*}}^3}{r\Omega} \right)\right],
\end{align}
with the constants $a_i^{h_{\ell m}}$ and  $b_i^{h_{\ell m}}$ chosen such that the EOB modes agree with NR modes at the point where the
inspiral-plunge modes are matched to the merger-RD modes (see Eq.~(\ref{EOB:IMR}) above and Ref.~\cite{Pompiliv5}).

In the initial {\tt SEOBNR} models, the NQC corrections were included in the energy flux (and RR forces) through Eq.~(\ref{eq:fluxmodes})
via an iterative procedure or fits (e.g., see Refs.~\cite{Pan:2011gk,Taracchini:2013rva}). However, starting from the {\tt SEOBNRv4} model
~\cite{Ossokine:2020kjp}, the inspiral-plunge modes in Eq.~(\ref{eq:fluxmodes}) only contain the factorized modes. The NQC corrections
are included only in the gravitational
polarization modes. The initial {\tt TEOBResumS} models also included the NQC corrections in the energy flux, but then in subsequent
versions the energy flux did not take them in. Recently,
the {\tt TEOBResumS} model of Ref.~\cite{Riemenschneider:2021ppj} incorporates fits to the NQC corrections in the energy
flux (and RR force) through Eq.~(\ref{eq:fluxmodes}). Finally, in the {\tt SEOBNRv5} waveform model used in this paper, the NQC corrections do
not enter the RR forces. As we shall discuss below, in the nonspinning case, the calibration to 2GSF results improves the EOB energy
fluxes considerably, thus the NQC corrections play a subdominant role. Furthermore, the inclusion of higher-order PN spin terms in the gravitational EOB modes,
which were absent in the previous {\tt SEOBNRv4} model, reduces the disagreement between EOB and NR fluxes and pushes it mostly to the very late inspiral,
where the effective test-body motion is almost unaffected by dissipative effects (see for details Ref.~\cite{Pompiliv5}).

\section{Basics of the gravitational self-force approach}\label{sec:GSF}

The development of the GSF approach has traditionally been driven by the need to model GW
emission from extreme-mass-ratio inspirals.  This approach expands the
metric of the binary around the metric of the primary in powers of $\mr = m_2/m_1$
(see Eq.~(\ref{eq:twotimescale-ansatz}) below).  It is well known~\cite{Hinderer:2008dm}
 that in order to get a waveform phase error that scales as $\mathcal{O}(\mr)$, the
expansion of the metric perturbation must be carried through
$\mathcal{O}(\mr^2)$ (i.e., at 2GSF).  That is to
say, the waveform phase error scales as $\mathcal{O}(\mr^0)$ if
second-order (in the mass-ratio) corrections are not included in the
metric.

Practical 2GSF calculations have recently been carried out using a
two-timescale framework~\cite{Miller:2020bft}.  Within this approach
the (multipolar) flux for a quasicircular, nonspinning binary was recently
computed~\cite{Warburton:2021kwk}.  This has since been combined with
a calculation of the binding energy \cite{Pound:2019lzj} to compute
the associated inspiral dynamics and waveforms \cite{Wardell:2021fyy}.
Important additional details are available in Ref.~\cite{Albertini:2022rfe}.
Here we will briefly review the calculation of the 2GSF flux.

Restricting to quasicircular orbits, we expand the metric of the binary as
\begin{align}\label{eq:twotimescale-ansatz}
	g_{\alpha\beta} + \sum^{\infty}_{m=-\infty}\left[\mr h^{1,m}_{\alpha\beta}(\Omega) + \mr^2 h^{2,m}_{\alpha\beta}(\Omega) \right]e^{-im\phi} + \mathcal{O}(\mr^3),
\end{align}
where $g_{\alpha\beta}$ is the Schwarzschild metric of the primary and
$\phi$ is the orbital azimuthal angle of the secondary.  The metric
amplitudes, $h^{n,m}_{\alpha\beta}$, depend on the binary's slowly
evolving orbital frequency $\Omega \equiv d\phi/dt$\footnote{The
  metric amplitudes also depend on small corrections to the primary's
  mass and spin which evolve due to absorption of gravitational
  radiation. The magnitude of these corrections is very small~\cite{Albertini:2022rfe}, so we
  ignore them in this work. Neglecting these effects, particularly the presence of a small but nonzero spin, is also consistent with the EOB model.}.  During the
inspiral (i.e., sufficiently far from the ISCO), the orbital frequency, and thus the metric amplitudes,
evolve on the slow RR timescale $t_{\rm RR} \sim
1/(\mr\Omega)$, whereas $\phi$ evolves on the fast orbital timescale
$t_{\rm orb}\sim1/\Omega$.  The two-timescale framework treats $t_{\rm
  RR}$ and $t_{\rm orb}$ as independent. By substituting Eq.~\eqref{eq:twotimescale-ansatz}
into the Einstein field
equations, we can split them into (i) a set of Fourier-domain partial differential
equations for the amplitudes at fixed $\Omega$, and (ii) evolution
equations that determine $\Omega$ and $\phi$ as functions of time.
Decomposing the metric perturbation onto a basis of tensor spherical
harmonics and working in the Lorenz gauge
\cite{Barack:2005nr,Barack:2007tm}, we arrive at the field equations
which can be found explicitly in Ref.~\cite{Miller:2020bft}.
Constructing the second-order source, applying appropriate boundary
conditions, and integrating the field equations required the
development of a raft of new techniques and codes~\cite{Pound:2014xva,Pound:2015wva,Miller:2016hjv,Akcay:2013wfa,Durkan:2022fvm,Wardell:2015ada}.

In our two-timescale scheme we assume the secondary follows a quasicircular orbit in which $\dot\Omega=O(\mr)\neq0$.
In order to satisfy the Einstein field equations through second order
in the mass ratio, we consistently account for the nonzero
$\dot{\Omega}$ everywhere that it appears (which is in numerous places
in the second-order field equation and second-order flux). However, the assumption $\dot{\Omega}\sim \mr$
breaks down near the ISCO, and the expansions based on that
assumption cause $\dot{\Omega}$ to unphysically diverge at the ISCO.
Due to the presence of $\dot{\Omega}$ terms, the 2GSF flux computed
using the above two-timescale expansion also diverges at the ISCO.
This non-physical divergence can be removed by transitioning to a new
expansion in the vicinity of the ISCO \cite{Compere:2021zfj}.  The
location where this transition occurs provides an estimate for where
the inspiral two-timescale expansion breaks down.
Reference~\cite{Albertini:2022rfe} estimated that, for small $\smr$, this
breakdown occurs around
\begin{equation}\label{eq:vbreak}
	 v_\Omega^{\rm break} = v_\Omega^{\rm ISCO} -0.052 \nu^{1/4},
\end{equation}
where $v_\Omega^{\rm ISCO} = 1/\sqrt{6} \approx 0.408$.
\begin{figure}
	\includegraphics[width=\columnwidth]{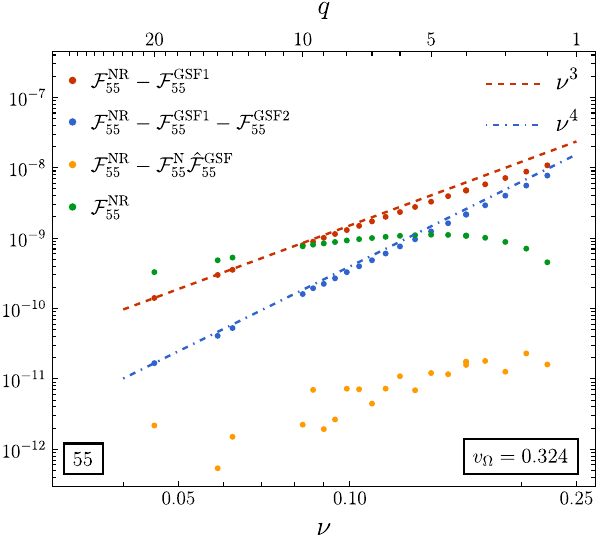}
	\caption{
		Comparison of the $(5,5)$-mode flux extracted from NR simulations and GSF calculations
at $r/GM \equiv 1/v^2_\Omega=9.5$ as function of the symmetric mass-ratio, $\smr$.
		After subtracting the 1GSF (resp.~2GSF) flux from the NR flux, we see the residual scales as $\smr^3$ ($\smr^4$), as expected.
		We further observe the significant improvement in the agreement with NR when using the re-expanded normalized GSF flux
defined in Eq.~\eqref{eq:FhatGSF_reexpand}.
		Similar results for the other mode fluxes considered in this work are given in Appendix~\ref{apdx:GSFvsNR}.
	}
	\label{fig:55-flux-scaling}
\end{figure}

The GSF energy
flux is calculated from the $(\ell, m)$ modes of $\mr
h^{1,m}_{\alpha\beta} + \mr^2 h^{2,m}_{\alpha\beta}$ at null infinity~\cite{Akcay:2010dx}.
With the polarizations expanded in -2 spin-weighted
spherical harmonics (see Eq.~\eqref{eq:waveform_harmonics}), the
flux is given by
\begin{align}
	\flux^{\rm GSF\mr}_{\ell m} =\lim_{d_L\to\infty} |\dot{h}_{\ell m}|^2 d_L^2/(16\pi).
\end{align}
Defining $y= (m_1 \Omega)^{2/3}$ we can write the flux as an expansion
in $\mr$ at fixed $y$ as
\begin{align}
	\flux^{\rm GSF\mr}_{\ell m}(\mr,y) = \mr^2 \flux^{\rm GSF1\mr}_{\ell m}(y) + \mr^3 \flux^{\rm GSF2\mr}_{\ell m}(y) + \mathcal{O}(\mr^4).
\end{align}
The symmetry of the physical binary system under the interchange of the labels $m_1\leftrightarrow m_2$ suggests that,
for comparable-mass binaries, it is natural to re-expand in the symmetric mass-ratio $\nu$ at fixed total mass $M$.
This is known to improve agreement of perturbative results with NR simulations of comparable-mass binaries~\cite{LeTiec:2011bk,vandeMeent:2020xgc}
and also it is natural for comparing with PN and EOB models.
Defining $x=(M\Omega)^{2/3} = v_\Omega^2$ and using
\begin{align}
	\mr &= \smr + 2\smr^2 +\mathcal{O}(\smr^3),\\
	y &= x(1-2/3\smr) + \mathcal{O}(\smr^2)\text{, and} \\
	m_1 &= M - m_2\smr + \mathcal{O}(\smr^2),
\end{align}
we can re-expand the flux as
\begin{align}
	\flux^{\rm GSF\smr}_{\ell m}(\smr,x) = \smr^2 \flux^{\rm GSF1\smr}_{\ell m}(x) + \smr^3 \flux^{\rm GSF2\smr}_{\ell m}(x) + \mathcal{O}(\mr^4),
\end{align}
where
\begin{subequations}
	\begin{align}
	\flux^{\rm GSF1\smr}_{\ell m}(x) &= \flux^{\rm GSF1\mr}_{\ell m}(x), \\
	\flux^{\rm GSF2\smr}_{\ell m}(x) &=  \flux^{\rm GSF2\mr}_{\ell m}(x)+ 4\flux^{\rm GSF1\mr}_{\ell m}(x) - x \d{\flux^{\rm GSF1\mr}_{\ell m}(x)}{x}.
\end{align}
\end{subequations}
Hereafter we shall use $\flux^{\rm GSF}_{\ell m} \equiv \flux^{\rm GSF\smr}_{\ell m}$. For interfacing with the EOB model, it is useful to define a re-expanded  (Newtonian-)normalized flux
\begin{equation}\label{eq:FhatGSF_reexpand}
	\frac{\flux^{\rm GSF}_{\ell m}}{\flux^{\rm N}_{\ell m}} = \hat{\flux}^{\rm GSF1}_{\ell m} + \smr \hat{\flux}^{\rm GSF2}_{\ell m} +\mathcal{O}(\smr^2),
\end{equation}
with
\begin{equation}
	\flux^{\rm N}_{\ell m} = d_L^2 \frac{(mM\Omega)^2}{8\pi}	\abs{h_{\ell m}^{\rm N}}^2.
\end{equation}
We hereafter define $\hat{\flux}^{\rm GSF}_{\ell m} \equiv \hat{\flux}^{\rm GSF1}_{\ell m} + \smr \hat{\flux}^{\rm GSF2}_{\ell m}$.
Using the re-expanded normalized flux also results in a significant improvement in the agreement between the GSF and NR fluxes for all modes other than the (2,2) mode. 
Presumably, this is due to this re-summed flux having the correct leading Newtonian behavior for all mass-ratios.

The first calculation of the 2GSF flux was presented in
Ref.~\cite{Warburton:2021kwk}, where remarkable agreement was found
between the total GSF flux (first plus second order) and the flux
computed from NR simulations of comparable-mass binaries.  Figure 4 of
Ref.~\cite{Warburton:2021kwk} showed that for the (2,2) and (3,3) mode the difference
between the NR and GSF flux scaled as $\mathcal{O}(\smr^4)$, as
expected, over mass ratios ranging from 10:1 to 1:1.  In
Fig.~\ref{fig:55-flux-scaling} we show that this scaling also holds for the
(5,5) mode across a wider range of mass ratios from 20:1 to 3:1.  In
Appendix~\ref{apdx:GSFvsNR} we show similar plots for the other modes
considered in this work.  This excellent agreement with NR gives us
further confidence that the calculated 2GSF flux is capturing all
contributions to the flux through $\mathcal{O}(\smr^3)$.  For this
work we also computed the 2GSF flux at many more orbital frequencies
than were previously presented in Ref.~\cite{Warburton:2021kwk}.

Figure~\ref{fig:55-flux-scaling} (and the related plots in
Appendix~\ref{apdx:GSFvsNR}) also shows the improvement gained by
using the re-expanded normalized GSF flux. We see that factoring out
the leading Newtonian behaviour brings the GSF flux much closer to
the NR flux, especially in the case of comparable masses. This factorization is
also a key part of the resummation of the EOB flux as described in
Sec.~\ref{sec:EOB}. This further motivates us to incorporate the new
2GSF flux information in the EOB flux.

\section{Matching EOB and GSF multipolar fluxes}

In order to incorporate information from the 2GSF flux into the EOB
flux, we need to compare the two in a gauge-invariant manner. In both
cases, the energy flux is decomposed into $-2$ spin-weighted spherical
modes. Consequently, we can compare the $(\ell, m)$-mode fluxes individually at a fixed value of the orbital frequency
$M\Omega$. By its nature, the GSF result is given as an expansion in
powers of $\smr$. We need to do the same with the EOB energy flux
modes $\flux_{\ell m}^{\rm EOB}$. Combining Eqs.~\eqref{eq:hfactors} and
$\eqref{eq:fluxmodes}$ gives
\begin{equation}\label{eq:Emodefact}
		\flux_{\ell m}^{\rm EOB}= d_L^2\frac{(m M\Omega)^2}{8\pi}
		\abs{h_{\ell m}^{\rm N}}^2
		\abs{\hat{S}_{\ell m}}^2
		\abs{T_{\ell m}}^2
		\abs{\rho_{\ell m}}^{2\ell}.
\end{equation}
Instead of expanding the full flux, it is more convenient to expand the flux normalized by its leading Newtonian contribution,
\begin{equation}\label{eq:normflux}
	\hat{\flux}_{\ell m}^{\rm EOB} = \frac{\flux_{\ell m}^{\rm EOB}}{\flux^{\rm N}_{\ell m}}.
\end{equation}
This allows us to preserve the nonpolynomial dependence of
$\flux^{\rm N}_{\ell m}$ on $\smr$ in the final EOB result, and ensure
its correct behaviour in the equal-mass limit. One potential
complication in doing this is that $\flux^{\rm N}_{\ell m}$ is not
written directly in terms of $\Omega$, but instead depends indirectly
on $\Omega$ through $v_\phi$. However, expanding the dependence of
$v_\phi$ in powers of $\smr$ and $v_\Omega$ (see Eqs.~(\ref{vp}) and (\ref{vO})), we find that
\begin{align}
	v_\phi &= v_\Omega + \mathcal{O}(\smr^2 v_\Omega^{13}).
\end{align}
Since in this comparison we are only interested in next-to-leading order corrections in $\smr$, we can thus safely replace $v_\phi$ by $v_\Omega$ everywhere in the expansion.

Next, we need to expand the individual factors in Eq.~\eqref{eq:normflux} in powers of $\smr$ at fixed values of $\Omega$.  Starting with the effective source $\hat{S}_{\ell m}$ given in Eq.~\eqref{eq:Slm}, we write
\begin{equation}
		\hat{S}_{\ell m}	=  \hat{S}_{\ell m}^{(0)} +\smr \hat{S}_{\ell m}^{(1)}  + \mathcal{O}(\smr^2).
\end{equation}
At leading order, this is simply given by the well-known test-body result
\begin{equation}
	\hat{S}_{\ell m}^{(0)}= \left\{\begin{aligned}
		&M\frac{1-2 v_\Omega^2}{\sqrt{1-3 v_\Omega^2}} , &&\text{$\ell+m$ is even},\\
		&M\frac{1}{\sqrt{1-3 v_\Omega^2}}, &&\text{$\ell+m$ is odd},
	\end{aligned} \right.
\end{equation}
which is reproduced exactly by the EOB Hamiltonian (by
construction). At next-to-leading order, the effective source for
quasicircular inspirals depends on the linear-in-$\smr$ correction
to the EOB $A$ potential. In principle, this contribution depends on
the exact details of the implementation of the A-potential in the {\tt SEOBNRv5HM}
model, including any calibration of $a_6$ to NR results. (This is one of the routes through
which the calibration of the A-potential becomes degenerate with
calibration of the energy flux in the EOB RR force.) However, through use of the first law of
binary mechanics \cite{LeTiec:2011ab}, it is possible to directly
compute the linear-in-$\smr$ correction to the $A$ potential
\cite{LeTiec:2011dp,Barausse:2011dq} in terms of the Detweiler
redshift invariant $z^{(1)}$ \cite{Detweiler:2008ft}, the exact value
of which can be computed numerically in the GSF context~\cite{Barausse:2011dq, Akcay:2012ea, vandeMeent:2015lxa}. This gives
\begin{equation}
\hat{S}_{\ell m}^{(1)}= \left\{\begin{aligned}
& H_{\rm eff}^{(1)} , &&\text{$\ell+m$ is even}\\
& L_{\rm eff}^{(1)}, &&\text{$\ell+m$ is odd},
\end{aligned} \right.
\end{equation}
with
\begin{subequations}
	\begin{align}
\frac{H_{\rm eff}^{(1)}}{M} 	&=
\frac{1}{2} z^{(1)}(v_\Omega) - \frac{1}{6} v_\Omega {z^{(1)}}'(v_\Omega)+\sqrt{1-3 v_\Omega^2}\\
&\quad+\frac{x(7-24 x)}{6(1-3 v_\Omega^2)^{3/2}}+\left(\frac{1-2 v_\Omega^2}{\sqrt{1-3 v_\Omega^2}} -1\right)^2-1, \notag\\
\frac{L_{\rm eff}^{(1)}}{M}	&= -\frac{1}{6v_\Omega} {z^{(1)}}'(v_\Omega)+ \frac{4-15 x}{6(1-3 v_\Omega^2)^{3/2}}.
\end{align}
\end{subequations}
We use interpolated data for the redshift $z^{(1)}$ generated with the code~\cite{vandeMeent:2015lxa} from a previous work~\cite{Antonelli:2019fmq}. This ensures that the matching procedure will not depend on precise feature of the dynamics in \SEOBNRvVHM, in particular removing any possibly circular dependence on the calibration to NR.

The next step is to expand the tail term $T_{\ell m}$, given in Eq.~\eqref{eq:Tlm}. This has a hidden dependence on $\smr$ through $\hat{\Omega} = \Omega H_{\rm EOB}$. We start by expanding  $H_{\rm EOB}$:
\begin{equation}
 H_{\rm EOB} =  H_{\rm EOB}^{(0)}+ \smr  H_{\rm EOB}^{(1)} + \mathcal{O}(\smr^2),
\end{equation}
which straightforwardly gives
\begin{subequations}
	\begin{align}
H_{\rm EOB}^{(0)} & = M,\\
H_{\rm EOB}^{(1)} & = M\left(\frac{1-2 v_\Omega^2}{\sqrt{1-3 v_\Omega^2}} -1\right).
\end{align}
\end{subequations}
Following~\cite{Damour:2008gu}, we note that the modulus square of the tail term can be written as
\begin{equation}
\abs{T_{\ell m}}^2 = \frac{1}{(\ell!)^2} \frac{4\pi m \hat{\Omega}}{1-e^{-4\pi m\hat{\Omega}}}
\prod_{k=1}^{\ell}\left[ k^2 + (2m\hat{\Omega})^2 \right].
\end{equation}
We thus find the expansion of this term as
\begin{equation}
	\abs{T_{\ell m}}^2 = \mathcal{T}_{\ell m}^{(0)} + \smr \mathcal{T}_{\ell m}^{(1)} +\mathcal{O}(\smr^2),
\end{equation}
with
\begin{align}
	\mathcal{T}_{\ell m}^{(0)}  &= \frac{1}{(\ell!)^2} \frac{4\pi m M\Omega}{1-e^{-4\pi m M\Omega}}
	\prod_{k=1}^{\ell}\left[ k^2 + (2m M\Omega)^2 \right],
	\\
\intertext{and}
	\frac{\mathcal{T}_{\ell m}^{(1)}}{\mathcal{T}_{\ell m}^{(0)}}  &=
	\frac{H_{\rm EOB}^{(1)}}{M}
	\bigg[
	1
	+ \frac{4\pi m M\Omega}{1-e^{4\pi m M\Omega}}
	+ \sum_{j=1}^{\ell} \frac{ 4m (M\Omega)^2}{ j^2 + (2m M\Omega)^2}
	\bigg].
\end{align}

We now write the expansion of the final factor in Eq.~\eqref{eq:Emodefact} as
\begin{equation}
\rho_{\ell m} = \rho_{\ell m}^{(0)} + \smr \rho_{\ell m}^{(1)} +\mathcal{O}(\smr^2),
\end{equation}
and compare to the re-expanded normalized GSF flux  in Eq.~\eqref{eq:FhatGSF_reexpand}.
We can find the exact values of $\rho_{\ell m}^{(0)}$ and $\rho_{\ell m}^{(1)}$ in terms of the GSF flux by matching the two expressions for the normalized flux at fixed $\Omega$ order-by-order in $\smr$, yielding
\begin{align}\label{eq:matchrho0}
	\rho_{\ell m}^{(0),\rm GSF} &= \Biggl(\frac{\hat{\flux}^{\rm GSF1}_{\ell m}}{\mathcal{T}_{\ell m}^{(0)}\abs{\hat{S}_{\ell m}^{(0)}}^2 }\Biggr)^{1/(2\ell)},\\
\intertext{and}\label{eq:matchrho1}
	\rho_{\ell m}^{(1),\rm GSF} &= 	\frac{\rho_{\ell m}^{(0)}}{2\ell} \left(\frac{\hat{\flux}^{\rm GSF2}_{\ell m}}{\hat{\flux}^{\rm GSF1}_{\ell m}} -\frac{\mathcal{T}_{\ell m}^{(1)}}{\mathcal{T}_{\ell m}^{(0)}} -2\frac{\hat{S}_{\ell m}^{(1)}}{\hat{S}_{\ell m}^{(0)}}
	\right).
\end{align}

Equation \eqref{eq:matchrho0}, of course, matches the expression previously found in \cite{Damour:2008gu}. The expression for $\rho_{\ell m}^{(1),\rm GSF}$ is the new expression needed to incorporate the 2GSF flux into the EOB flux.


\begin{figure}[!tb]
	\includegraphics[width=\columnwidth]{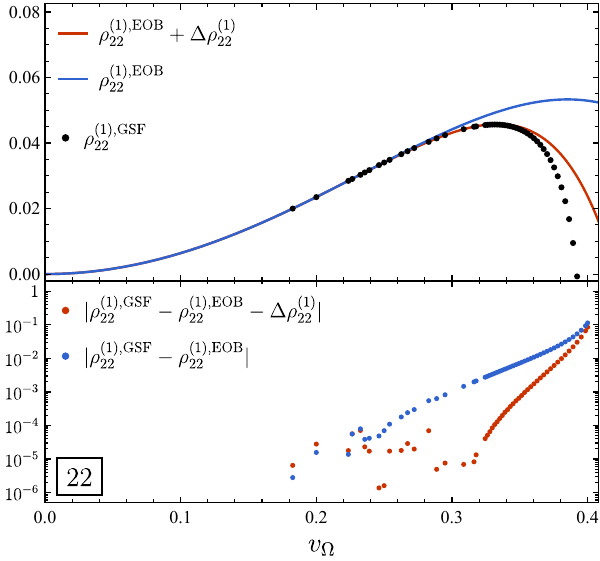}
	\caption{The top panel shows the numerical values of $\rho_{22}^{(1),\rm GSF}$ from applying Eq.~\eqref{eq:matchrho1} to the 2GSF fluxes, and the base EOB $\rho_{22}^{(1),\rm EOB}$ given by~\eqref{eq:rho1EOB22}. In addition, we show the corrected $\rho_{22}^{(1)}$ after adding the fitted correction~\eqref{eq:deltarho22}. The bottom panel shows the absolute difference between the GSF and EOB values with and without $\Delta\rho_{22}^{(1)}$.
	}
	\label{fig:rho22}
\end{figure}

\begin{figure*}[!p]
	\includegraphics[width=0.9\columnwidth]{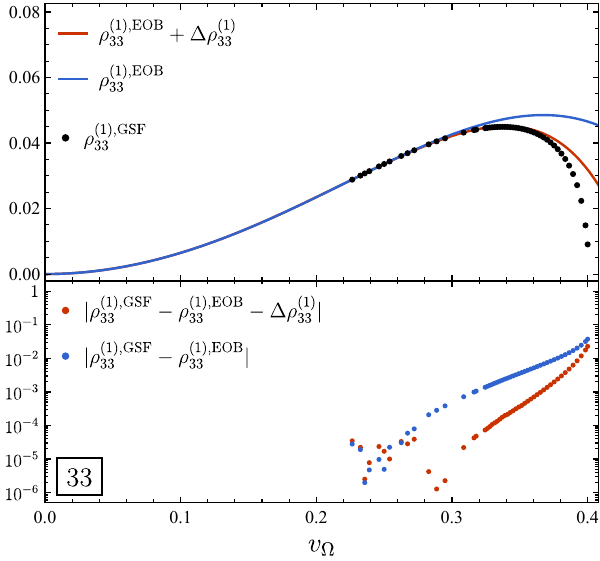}
	\includegraphics[width=0.9\columnwidth]{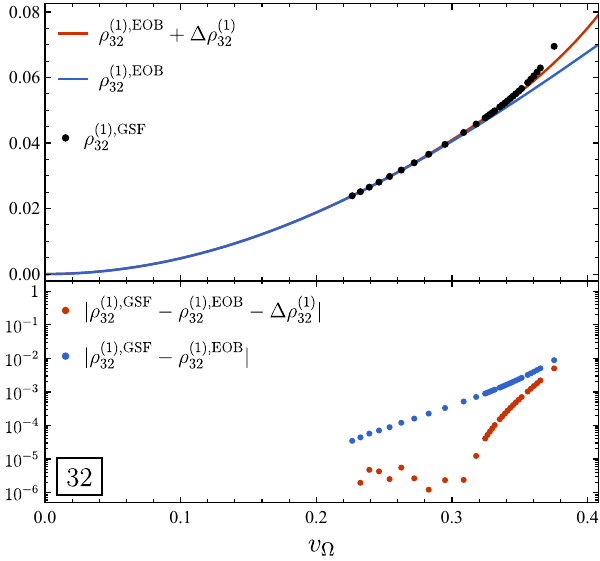}
	\includegraphics[width=0.9\columnwidth]{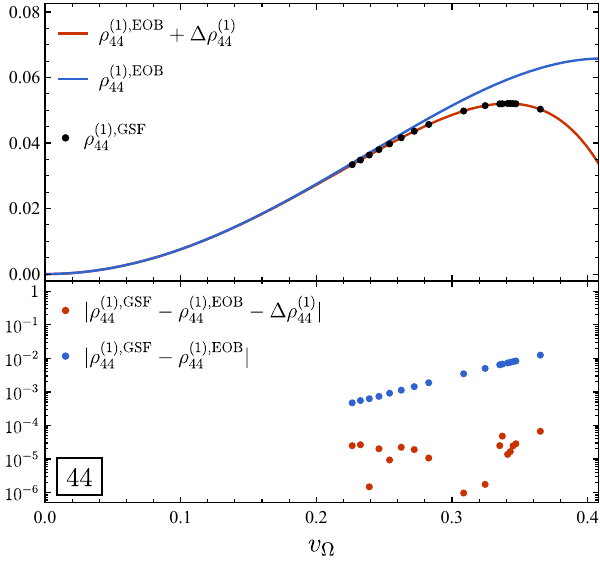}
	\includegraphics[width=0.9\columnwidth]{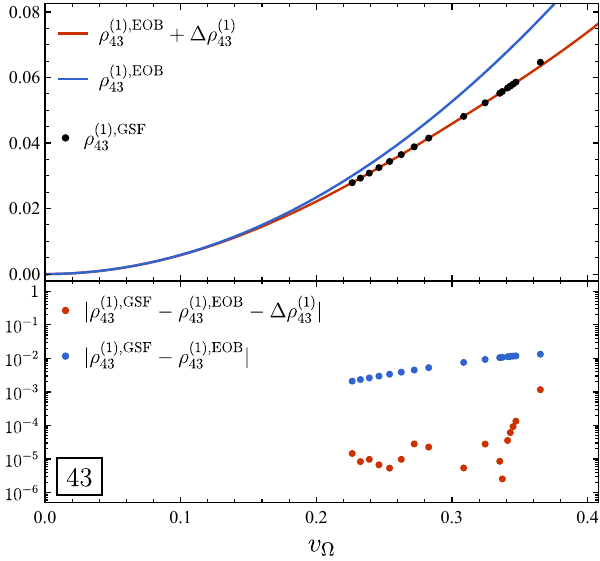}
	\includegraphics[width=0.9\columnwidth]{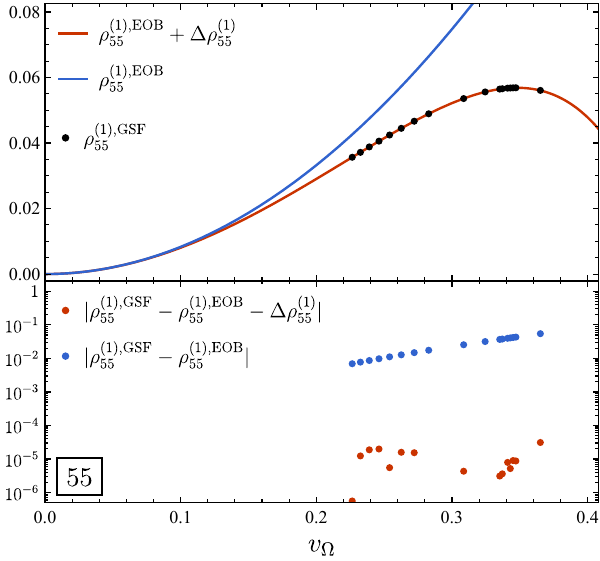}
	\includegraphics[width=0.9\columnwidth]{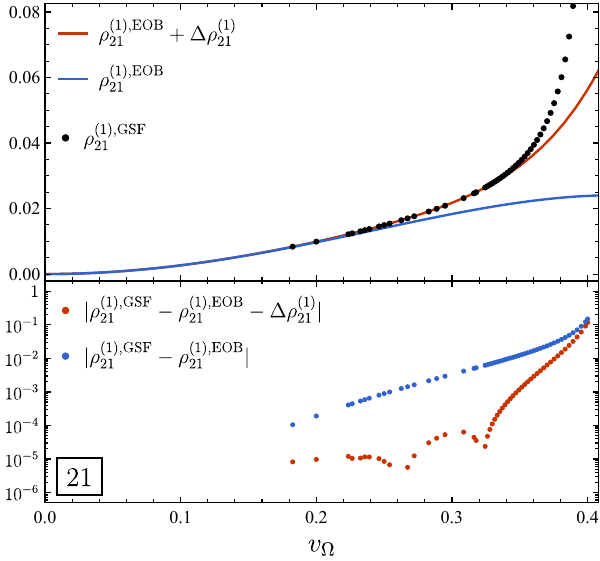}
	\caption{Same as in Fig.~\ref{fig:rho22}, but now for the (3,3), (3,2), (4,4), (4,3), (5,5), and (2,1) modes.
The panels on the left show the even parity modes, while the panels on the right show the odd parity modes. We note a different
behaviour of the even and odd parity 2GSF modes close to the divergence induced by the two-timescale expansion in the proximity of the ISCO.
	}
	\label{fig:multirho}
\end{figure*}

For including 2GSF information in the EOB mode amplitudes and
  energy flux, we focus on the 7 dominant $(\ell, m)$ modes that are
included in the {\tt SEOBNRv5HM} model. For these modes we first determine the
contributions to $\rho_{\ell m}^{(1)}$ already included in the EOB
waveform modes by expanding $\rho_{\ell m}$ in powers of
$\smr$,\footnote{Note that the \SEOBNR{v5HM} model does not include all PN
    information available at the time of writing (for details see Appendix~B
    of Ref.~\cite{Pompiliv5}).}
\begin{subequations}\label{eq:rho1EOB22}
\begin{align}
	\rho_{22}^{(1),\rm EOB} &=
\tfrac{55 }{84}v^2_\Omega
-\!\tfrac{33025 }{21168}v^4_\Omega
-\!\big[\tfrac{48993925}{9779616}-\tfrac{41 \pi ^2}{192}\big] v^6_\Omega,
\\
\rho_{21}^{(1),\rm EOB} &=
\frac{23}{84} v^2_\Omega
-\frac{10993}{14112} v^4_\Omega,
\\
\rho_{33}^{(1),\rm EOB} &=
\tfrac{2 }{3}v^2_\Omega
-\tfrac{1861}{990} v^4_\Omega
-\left[\tfrac{129509}{25740}-\tfrac{41 \pi ^2}{192}\right] v^6_\Omega,
\\
\rho_{32}^{(1),\rm EOB} &=
\frac{131}{270} v^2_\Omega
-\frac{617123}{1603800} v^4_\Omega,
\\
\rho_{44}^{(1),\rm EOB} &=
\frac{257}{330} v^2_\Omega
-\frac{5072887}{2202200} v^4_\Omega,
\\
\rho_{43}^{(1),\rm EOB} &= \frac{103}{176}  v^2_\Omega,
\\
\rho_{55}^{(1),\rm EOB} &= \frac{54}{65} v^2_\Omega.
\end{align}
\end{subequations}
We augment the $\rho_{\ell m}^{(1),\rm EOB}$ by adding an additional
polynomial $\Delta\rho_{\ell m}^{(1)}$ in $v^2_\Omega$ starting at the
lowest order in $v^2_\Omega$ not already included, no power higher than $v^{10}_\Omega$, and at most three terms. The
$\Delta\rho_{\ell m}^{(1)}$ are determined by fitting to the numerical
$ \rho_{\ell m}^{(1),\rm GSF}$ results. While these extra terms take
the form of higher-order PN terms, we emphasize that the goal here is
not to estimate the next-order terms in the PN series (which would in
general also contain $\log v_\Omega$ contributions). Instead, the goal
is to capture as much of the behaviour of the numerical $\rho_{\ell
  m}^{(1),\rm GSF}$ data as possible.

To see how this fit works in practice, let us focus on the case of the
$(2,2)$-mode shown in Fig.~\ref{fig:rho22}. There are two complicating
factors in performing the fit. The first is that the GSF data has
finite numerical accuracy. This causes issues in the weak-field
regime, where the EOB approximation is more accurate than the GSF
data, and we thus run the risk of overfitting the numerical noise. The
second complication is that the GSF data diverges at the Schwarzschild
ISCO at $v_\Omega=1/\sqrt{6}$, where the inspiral two-timescale expansion
breaks down (see Sec.~\ref{sec:GSF}). This feature is not physical and should not be reproduced
by the EOB flux.

As a result of these complications the residual after subtracting $\rho_{\ell m}^{(1),\rm EOB}$ from $\rho_{\ell m}^{(1),\rm GSF}$ has three main features (as visible in the lower panel of Fig.~~\ref{fig:rho22}):
	 In the low frequency regime the residual is (nearly) constant, indicating the numerical noise floor.
	 At high frequencies the residual shows a sharp increase due to the divergence at the ISCO.
	 In the middle the residual scales with a power law compatible with the lowest unknown PN orders.
	 Our goal is to fit this middle feature without digging into either source of systematic bias. 
	 Typical automated fitting procedures will fail at the latter, and adjusting them to avoid doing so will typically introduce more new adjustable parameters than are being fitted for in the first place. 
	 Consequently, the most practical approach is to manually adjust the fitting parameters until the middle feature completely disappears, and the residual is completely dominated by the systematic biases from either end of the spectrum.
  In the case of the $(2,2)$-mode this produces
\begin{subequations}
\begin{equation}\label{eq:deltarho22}
	\Delta\rho_{22}^{(1)} =
	21.2 v^8_\Omega
	-411v^{10}_\Omega.
\end{equation}

Repeating the process for the six remaining modes (shown in Fig.~\ref{fig:multirho}) yields the following fits:
\begin{align}
\Delta\rho_{21}^{(1)} &=
1.65  v^6_\Omega
+26.5 v^8_\Omega
+80   v^{10}_\Omega,
\\
\Delta\rho_{33}^{(1)} &=
12 		v^8_\Omega
-215 	v^{10}_\Omega,
\\
\Delta\rho_{32}^{(1)} &=
0.333  v^6_\Omega
-6.5 v^8_\Omega
+98   v^{10}_\Omega,
\\
\Delta\rho_{44}^{(1)} &=
-3.56  v^6_\Omega
+15.6 v^8_\Omega
-216   v^{10}_\Omega,
\\
\Delta\rho_{43}^{(1)} &=
-0.654  v^4_\Omega
-3.69 v^6_\Omega
+18.5   v^8_\Omega,
\\
\Delta\rho_{55}^{(1)} &=
-2.61  v^4_\Omega
+1.25 v^6_\Omega
-35.7   v^8_\Omega.
\end{align}
\end{subequations}

Thus, in the GSF-augmented EOB model, $\rho_{\ell m}$ in Eq.~\eqref{eq:flm} is replaced as
\begin{equation}
\rho_{\ell m} \mapsto \rho_{\ell m}+ \nu \Delta\rho_{\ell m}^{(1)},
\end{equation}
both when computing the EOB gravitational polarizations and RR
  force (taking $\Delta\rho_{\ell m}^{(1)}=0$ for modes for which it has not
been calculated, yet).

For two modes (the $(3,2)$ and $(4,3)$) these fits contain
higher PN terms than included in the corresponding $\rho_{\ell
  m}^{(0)}$ in previous \texttt{SEOBNR} models, which included terms up to $v^8_\Omega$ and $v^6_\Omega$ respectively. For the sake of
consistency, \SEOBNR{v5HM}~\cite{Pompiliv5} augments the corresponding $\rho_{\ell
  m}^{(0)}$-terms with terms at $v^{10}_\Omega$ and $v^8_\Omega$ using 1GSF
flux terms, which are known up to very high PN
order~\cite{Fujita:2012cm}.

\section{Impact of GSF information on the {\tt SEOBNRv5HM} model accuracy}\label{sec:results}

In this section we study the impact of including the 2GSF information in the energy flux and mode amplitudes on the overall
faithfulness of the \SEOBNR{v5HM} model developed in Ref.~\cite{Pompiliv5}.

\begin{figure}[!tb]
	\includegraphics[width=\columnwidth]{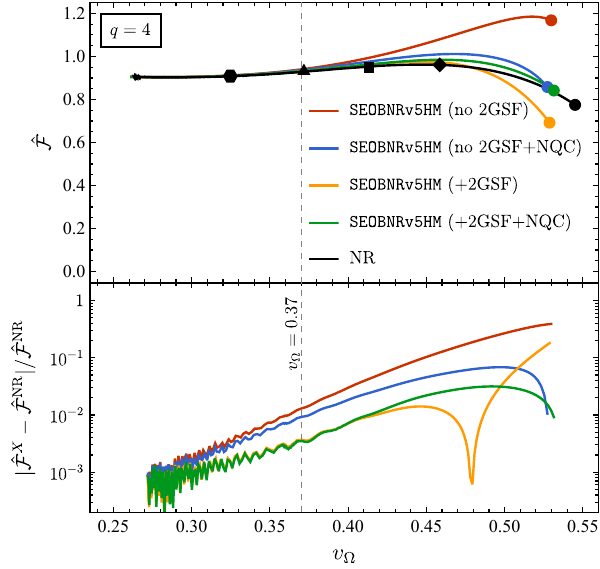}
	\caption{The normalized energy flux $\hat{\flux}$ compared between a quasicircular nonspinning NR simulation at mass-ratio $\lmr = 4$, and the {\tt SEOBNRv5HM} flux with and without 2GSF calibration. In addition we show what happens when the NQC corrections are included in the {\tt SEOBNRv5HM} flux. The top panel shows the full flux as a function of $v_\Omega$. The circles indicate the merger (peak of $|h_{22}^{\rm insp-plunge}|$), and the other markers indicate 1 (diamond), 2 (square), 4 (triangle), and 10 (hexagon) GW cycles before merger. The bottom panels show the relative difference between NR and the {\tt SEOBNRv5HM} fluxes. The vertical dashed line indicates the fixed frequency used for Fig.~\ref{fig:NRflux2}.
	}
	\label{fig:NRflux}
\end{figure}

\begin{figure}[!tb]
	\includegraphics[width=\columnwidth]{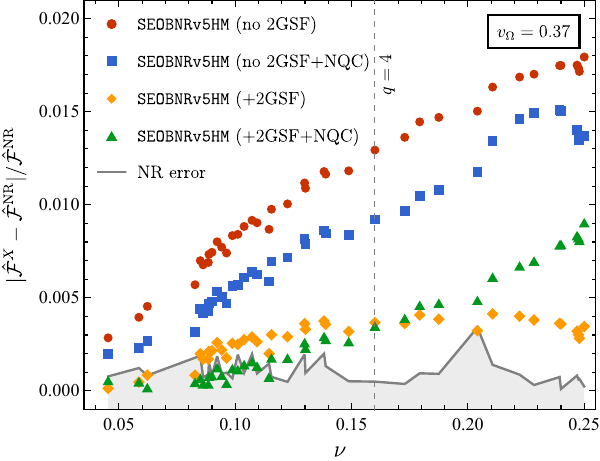}
	\caption{The relative difference at fixed frequency between the energy flux from NR and {\tt SEOBNRv5HM} fluxes with and without 2GSF and NQC correction for a range of mass ratios. The vertical line shows the mass ratio used in Fig.~\ref{fig:NRflux}.
		The shaded region at the bottom indicates an estimate of the uncertainty in the NR data obtained by varying $v_\Omega$ between $0.365$ and $0.375$.}
	\label{fig:NRflux2}
\end{figure}

We start with comparing the energy flux of the \SEOBNR{v5HM} model to
NR simulations from the SXS collaboration~\cite{Boyle:2019kee,Varma:2018sqd}. Details of the simulations used can be found in Appendix~\ref{app:nrsims}.
In Fig.~\ref{fig:NRflux} we compare the energy flux of an NR
simulation with mass-ratio $\lmr = m_1/m_2 = 4$ to the {\tt SEOBNRv5HM}
flux~\eqref{eq:EOBflux} with and without the 2GSF corrections. We see
that even at the modest mass-ratio, the 2GSF corrections improve the
agreement with the NR flux by a factor of a few across the frequencies
spanned. This improvement is much more substantial than that obtained
by including the NQCs in the energy flux~\cite{Pompiliv5}. Moreover, we see that adding the
NQCs to the flux on top of the 2GSF corrections leads to no
substantial improvement except in the last fraction of a GW cycle before merger.

To understand how the improvement of the {\tt SEOBNRv5HM} flux due to the 2GSF
corrections scales with mass-ratio, in Fig.~\ref{fig:NRflux2} we plot the same quantities as in Fig.~\ref{fig:NRflux}, but now for different NR simulations at varying mass-ratio and fixed value of $v_\Omega = 0.37$.
We again see that the 2GSF calibration improves the agreement with the NR flux by a factor of a few across the range
of sampled mass-ratios, and provides a much more substantial
improvement than merely including NQC corrections in the energy flux.
At low $\smr$, adding the NQC corrections on top of the 2GSF corrections provides an
additional small improvement, while near equal masses the NQC corrections actually make the agreement with NR slightly worse.
Naively, one might expect the relative error of the \SEOBNR{v5HM} flux with the 2GSF calibration to scale with $\smr^2$.
However, instead we see a relative error which is almost constant.
This suggests that insufficient accuracy in the $\rho_{\ell m}^{(0),\rm EOB}$ (i.e. the test-body flux) is the dominant source of error (see also~\cite{Albertini:2022dmc,Albertini:2022rfe}).
However, note that while $v_\Omega = 0.37$ is smaller than $v_\Omega^{\rm break}$~\eqref{eq:vbreak} for all mass-ratios, it is still close enough to the ISCO for corrections to the flux from the transition to plunge to be relevant.
Such contributions would lead to an almost flat relative error scaling as $\nu^{2/5}$ (see e.g. Fig.~\ref{fig:l2m2-flux-scaling-with-transition} in Appendix~\ref{apdx:GSFvsNR}).

\begin{figure*}[!tb]
	\centering
	\includegraphics[width=1\textwidth]{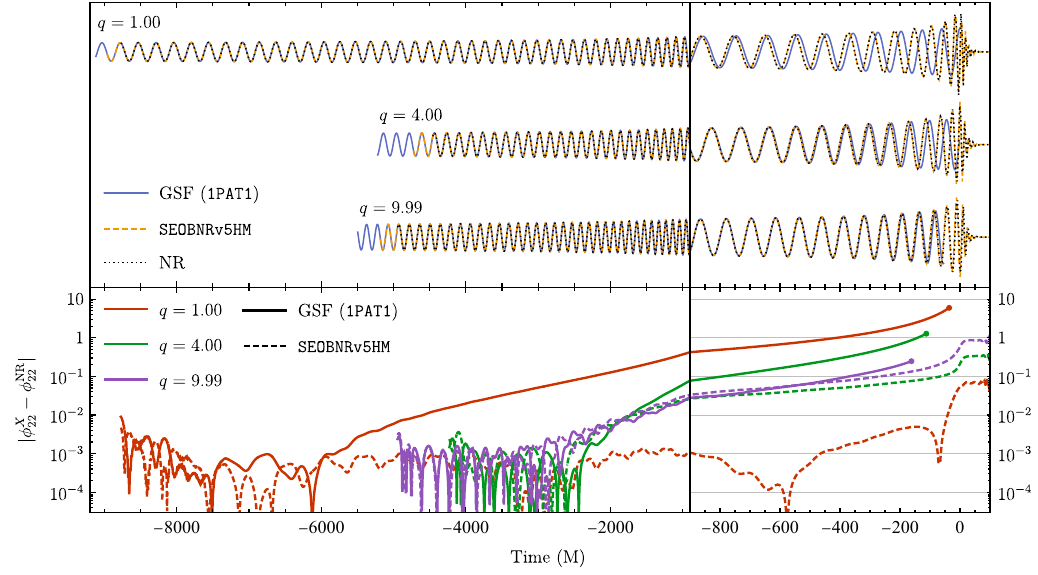}
	\caption{Top panel: Comparison of the (2,2)-mode waveforms from NR, the GSF \texttt{1PAT1} model and the \SEOBNR{v5HM} model at three different mass-ratios $q$. The waveforms at each mass-ratio are aligned at the start of the NR waveforms using the procedure described in Ref.~\cite{Pompiliv5}. The last $-900 M$ before the peak of the (2,2)-mode are shown magnified. Bottom panel: Dephasing of the \texttt{1PAT1} and \SEOBNR{v5HM} models relative to the NR waveforms. The GSF waveforms are truncated at $v_\Omega^{\rm break}$ \eqref{eq:vbreak} indicated by the dots. }
	\label{fig:WFplot}
\end{figure*}

\begin{figure}[!tb]
	\includegraphics[width=\columnwidth]{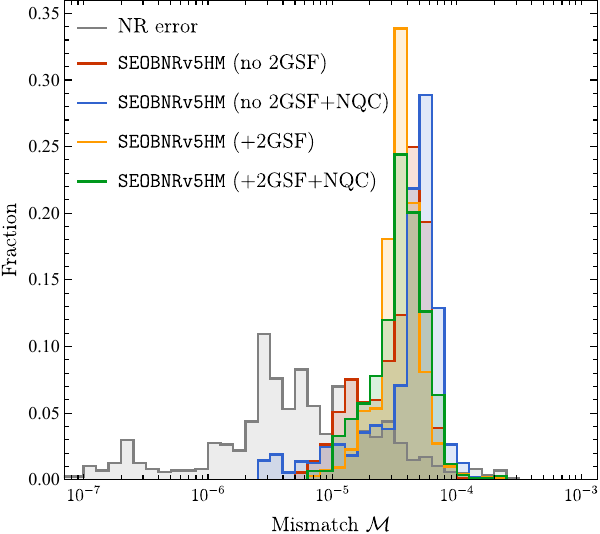}
	\caption{A histogram of the mismatches of NR versus {\tt SEOBNRv5HM} models with and without 2GSF and NQC corrections. As an indicator of the NR error, the mismatch of the highest resolution NR waveforms against the next higher resolution is shown in gray.}
	\label{fig:mismatchcal}
\end{figure}

It thus appears that the calibration of the {\tt SEOBNRv5HM} flux against 2GSF results
is successful in bringing the EOB flux closer to the NR
flux. Ultimately, the true measure of the model is the waveforms that
it produces. In Fig.~\ref{fig:WFplot} we compare the (2,2)-mode of the
\SEOBNR{v5HM} model (including 2GSF calibration) to waveforms from the
pure GSF-based waveform from Ref.~\cite{Wardell:2021fyy} (in its
\texttt{1PAT1} form)\footnote{The \texttt{1PAT1} GSF waveform
    model makes a number of approximations. A detailed discussion of
    the approximations used and domain of \texttt{1PAT1}'s validity
    can be found in Sec.~II of Ref.~\cite{Albertini:2022rfe}.}, and NR
waveforms at three different
mass ratios. The 2GSF waveforms are shown until the
binary velocity reaches $v_{\Omega}^{\rm break}$. For the first part, the waveforms are visually
indistinguishable, and only in the last $\sim 900 M$  before merger
we start to see differences, especially with the (inspiral
only!) 2GSF waveforms with mass ratios 4 and 1. Indeed, when we look at the dephasing in the
lower panel, the 2GSF waveforms for mass ratios 9.99 and 4 (1)  stay
below 0.1 (0.3) radians up to $\sim 900 M$ before
merger. Remarkably, the dephasing of the 2GSF waveform for mass
ratio 9.99 is still below 0.1 radians when the velocity reaches $v_{\Omega}^{\rm break}$,
highlighting the importance of including large mass-ratio corrections,
while for mass ratios 4 and 1, the dephasing reaches 1 radian and $\sim 6$
radians, respectively,  at $v_{\Omega}^{\rm break}$.
The dephasing of the \SEOBNR{v5HM} waveforms is shown throughout
the coalescence (i.e., during the inspiral, merger and ringdown stages)
and it is much smaller than that of the GSF waveforms. This is expected
since the \SEOBNR{v5HM} waveforms have been calibrated to NR simulations~\cite{Pompiliv5}.

To provide a more quantitative assessment of the impact of including the
2GSF information in the \SEOBNR{v5HM} model, we calculate the mismatch (or unfaithfulness)
between (2,2)-modes of the {\tt SEOBNRv5HM} waveforms and of a set of NR
waveforms with varying mass ratios using (e.g., see Ref.~\cite{Pompiliv5})
\begin{equation}
\mathcal{M} = 1-
	\max_{\delta\phi,\delta{}t} \frac{( h_{22}^{\rm NR}\vert h_{22}^{\rm EOB} )}{\sqrt{( h_{22}^{\rm NR}\vert h_{22}^{\rm NR} )( h_{22}^{\rm EOB}\vert h_{22}^{\rm EOB} )}},
\end{equation}
where we maximize over the relative shift in phase ($\delta\phi$) and time ($\delta{t}$) between the two waveforms, while $(\cdot|\cdot)$ denotes the noise weighted inner product~\cite{Finn:1992xs,Sathyaprakash:1991mt} given by
\begin{equation}
	\left(h_1 \mid h_2\right) \equiv 4 \operatorname{Re} \left [\int_{f_l}^{f_h} \frac{\tilde{h}_1(f) \tilde{h}_2^*(f)}{S_n(f)} \mathrm{d} f \right ] ,
\end{equation}
where $S_n(f)$  is the one-sided power
spectral density (PSD) of the detector noise, which we assume to be the design zero-detuned high-power
noise of Advanced LIGO~\cite{Barsotti:2018}.

For each NR simulation, we calculate the mismatches for a range of
total masses between $10 M_{\odot}$ and $290
M_{\odot}$. In Fig.~\ref{fig:mismatchcal}, we show a histogram of the
mismatches for the three cases: the \SEOBNR{v5HM} model without
2GSF corrections, the \SEOBNR{v5HM} model with 2GSF corrections in the RR force (i.e., in the energy flux) and
polarization modes, and the latter with the addition of NQC corrections in the
RR force.  Each waveform model is calibrated to NR by tuning the two calibration
parameters introduced in Sec.~\ref{sec:EOB}: $a_6$, which appears in the Hamiltonian, and $\Delta
t^{22}_{\text{ISCO}}$, which determines the time at which the inspiral-plunge waveform
is matched to the merger-ringdown one (for details see Ref.~\cite{Pompiliv5}). We stress that the
NR calibration is done by demanding that the {\tt SEOBNRv5} (2,2) mode has mismatches
with NR below $10^{-3}$ throughout the inspiral, merger and ringdown stages.

After calibration to the NR simulations, the histograms of the three models in Fig.~\ref{fig:mismatchcal} are very
  similar.  To gain insight into this, it is instructive to compare
  the mismatches between the models and NR to the NR error.
  Generally, there are several contributions to the NR-error budget,
  including truncation error and error in extrapolating the waveforms
  to infinity.\footnote{It should be noted that there are other
    sources of error in the NR simulations, in particular due to
    residual spin and residual eccentricity.  For the configurations
    considered here, these effects are subdominant.} For mismatch studies,
  the former is more dominant~\cite{Boyle:2019kee}, so
  we compute the mismatch between the highest and second highest NR
  resolution as a conservative measure of the NR error.  From
  Fig.~\ref{fig:mismatchcal} one can see that the mismatches between
  the models and NR are close to the NR error.  This suggests that
  part of the reason for minor differences between the models is that they
  are hitting the limits due to NR error.

\begin{figure}
	\includegraphics[width=\columnwidth]{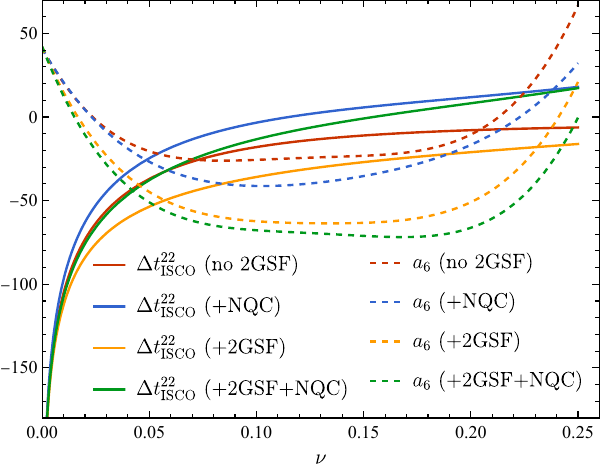}
	\caption{The calibration parameters $a_6$ and $\Delta t^{22}_{\text{ISCO}}$ as a function of the symmetric mass-ratio $\smr$ for \SEOBNR{v5HM} models with and without 2GSF corrections to the RR force (or in the energy flux) and the gravitational modes.}
	\label{fig:calibration}
\end{figure}

However, a potentially more important factor is that there is a large
degree of degeneracy in the mismatch between changes in the RR force and
changes in the Hamiltonian controlling the conservative dynamics (see Eqs.~\eqref{eq:EOB-EOMs} and~\eqref{eq:matchrho1}).
Consequently, the calibration of the Hamiltonian (through the calibration of the
waveform modes) can largely compensate for imperfections in the dissipative sector of the
EOB approach. In Fig.~\ref{fig:calibration}, we see the values of the two main
calibration parameters, $a_6$ and $\Delta t^{22}_{\text{ISCO}}$ of the
\SEOBNR{v5HM} models with and without 2GSF corrections in the RR force and polarization
modes. The presence of the 2GSF corrections has clear impact on the
calibration coefficients. This implies that the two calibrated models
have somewhat different dynamics; however, those differences do not
lead to appreciable differences in the corresponding waveforms, as can be seen in Fig.~\ref{fig:mismatchcal}.

\begin{figure}[!tb]
	\includegraphics[width=\columnwidth]{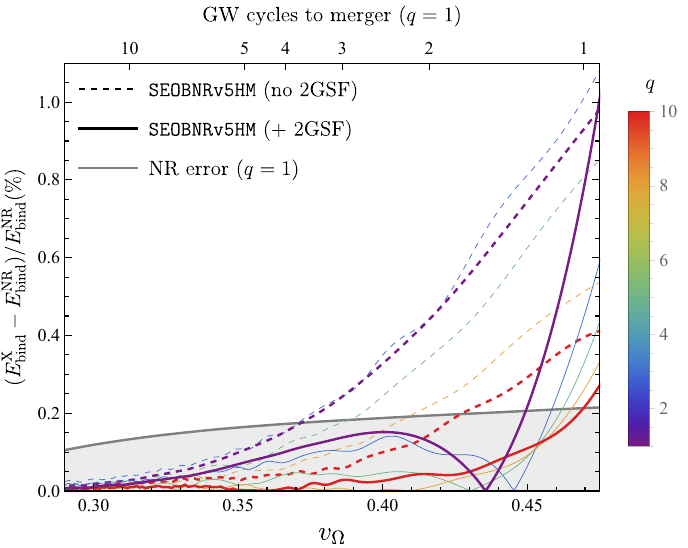}
	\caption{Relative difference between the binding energy $E_{\rm bind}^{\rm NR}$ inferred from NR simulations, and $E_{\rm bind}^{\rm EOB}$ from the \SEOBNR{v5HM}  models with and without 2GSF corrections. The shaded area shows the estimated error on the NR binding energy in the case $q=1$, which is taken as representative for the general NR error. The ticks on the top $x$-axis show the number of GW cycles before merger for the $q=1$ NR simulation.
}
	\label{fig:bindingenergy}
\end{figure}

\begin{figure}[!tb]
	\includegraphics[width=\columnwidth]{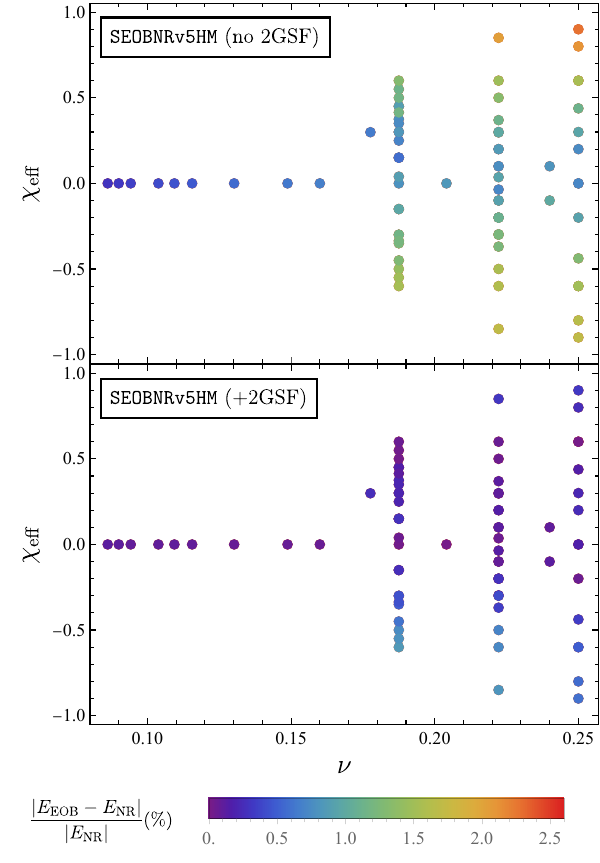}
	\caption{Comparison of the {\tt SEOBNRv5HM} binding energy to NR data for spin-aligned waveforms at a fixed value $v_\Omega=0.447$. The top (bottom) panel shows the binding energy from the \SEOBNR{v5HM} model with (without) 2GSF information.}
	\label{fig:alignedbindE}
\end{figure}

Since the calibration parameter $a_6$ controls part of the EOB $A$-potential, the two {\tt SEOBNRv5HM} models with and without 2GSF information
have different Hamiltonians, and therefore differ in their binding energy. The latter is given by
\begin{equation}
\label{eq:EOBbinding}
{E_{\rm bind}^{\rm EOB}}= H_{\rm EOB} - M.
\end{equation}
In Fig.~\ref{fig:bindingenergy} we compare the {\tt SEOBNRv5HM}
  binding energy to the one extracted from NR simulations from
Ref.~\cite{Ossokine:2017dge}. The {\tt SEOBNRv5HM} with 2GSF corrections reproduces
the NR binding energy much more faithfully, staying within the NR
error estimates until roughly one GW cycle before merger.
This improvement persists even when modeling aligned-spin binaries despite
only adding 2GSF corrections to the nonspinning part of the waveform
and RR force. In Fig.~\ref{fig:alignedbindE}, we compare the
{\tt SEOBNRv5HM} binding energy for models with and without the 2GSF information
to that of a set of spin-aligned NR waveforms~\cite{Ossokine:2017dge} at a fixed value $v_\Omega = 0.447$.
Without the 2GSF calibration, the binding energy can be off by as much as 2.5\% especially
at high values of the effective spin $\chi_{\rm eff}=(\chi_1 m_1 + \chi_2 m_2)/M$. However, with
the 2GSF calibration, we find deviations from the NR
binding energy to be at the sub-percent level, with an average difference
of just $0.24\%$.

\section{Discussion}
In this paper, we have enhanced the accuracy of the (factorized) gravitational modes used in the
\texttt{SEORBNRv5} models of Refs.~\cite{Khalilv5,Pompiliv5,Ramos-Buadesv5}
by calibrating them to nonspinning, quasi-circular 2GSF multipolar data of Ref.~\cite{Warburton:2021kwk}. This calibration affects also the
EOB radiation reaction~(RR) force driving the dynamical evolution of the binary black holes in the \texttt{SEORBNRv5} model.

By direct comparison of the energy flux in the \texttt{SEORBNRv5HM} model to that extracted from NR
simulations, we have confirmed in Figs.~\ref{fig:NRflux} and \ref{fig:NRflux2} that the 2GSF calibration of the flux leads to a significant improvement
in the faithfulness of the \texttt{SEORBNRv5HM} flux. In particular, the improvement seems to make the inclusion
of NQC corrections in the RR force subdominant, and limited to the very late inspiral (plunge),
where the effective test-body motion is almost unaffected by dissipative effects.

Furthermore, when looking at the mismatches between the
\texttt{SEORBNRv5HM} and NR waveforms in Fig.~\ref{fig:mismatchcal}, the
inclusion of the 2GSF calibration seems to only have a marginal impact
on the waveform mismatches after calibration to NR. If anything, this is a
testament to the effectiveness of the \texttt{SEORBNRv5} Hamiltonian's
calibration to NR, which is obtained by demanding that the
mismatches of the \texttt{SEORBNRv5} inspiral-merger-ringdown (2,2) modes are below $10^{-3}$.
Since the waveforms are computed using the EOB equations of motion,
which depend on the conservative and dissipative dynamics, the mismatches have
significant degeneracy between the calibration terms in the
EOB Hamiltonian and in the RR force (notably the 2GSF terms in the energy
flux). This is one reason why the flux calibration
terms that we have added to the EOB flux could not have been added through
the NR calibration performed in Ref.~\cite{Pompiliv5}\footnote{One could explore
in the future the possibility of calibrating directly the Hamiltonian from the
binding energy extracted from NR simulations instead of doing it indirectly
using the waveforms~\cite{Ossokine:2017dge}. However, this procedure would require the computation of the binding energy
for the entire set of $442$ aligned-spin SXS NR waveforms used to calibrate the {\tt SEOBNRv5} model.}.
This degeneracy also means that
calibrating \SEOBNR{v5HM} to NR with and without the 2GSF calibration of the
flux leads to a different EOB Hamiltonian. The Hamiltonian
itself however is supposed to correspond to a gauge invariant
observable of the binary, the binding energy. Comparing the EOB
binding energy to results extracted from NR simulations in Fig.~\ref{fig:bindingenergy}, we find that
the Hamiltonian calibrated with the 2GSF information included
reproduces the NR binding energy much more faithfully than the
Hamiltonian calibrated without. This is a significant consistency test
of the \SEOBNR{v5HM} model, and one that extends to binary BHs with spins,
as well (see Fig.~\ref{fig:alignedbindE}). So, while adding the 2GSF information to the \SEOBNR{v5HM} model does not
necessarily improve the faithfulness of the corresponding waveforms in the
regime where they are calibrated to NR, it does improve the overall
consistency and naturalness of the model. This gives us greater confidence that the
{\tt SEOBNRv5HM} model will remain (somewhat) faithful to NR when extrapolated beyond the
calibration region, in particular for higher mass ratios.

In this work we focused on adding 2GSF corrections to the nonspinning
sector of the {\tt SEOBNRv5HM} waveforms.  However, numerical results are also available
for corrections to the 2GSF flux linear in either the primary or secondary
spin~ \cite{Warburton:2021kwk,Akcay:2019bvk,Mathews:2021rod}.  In
principle, the matching procedure employed here can also be used to
calibrate the {\tt SEOBNRv5HM} modes to these data. We will leave the
implementation of this to future work.

A limiting factor in this work has been that the 2GSF multipolar flux data we used do
not include corrections due to the transition from inspiral to plunge,
causing it to diverge at the ISCO. This has limited our
ability to calibrate the RR force and gravitational modes in
the strong-field regime. Inclusion of these terms could lead to
further improvements of our results, and will be addressed once
new 2GSF data becomes available.

\section*{Acknowledgments}

The authors thank H\'ector Estell\'es, Mohammed Khalil, and Antoni Ramos-Buades for useful discussions.
MvdM is supported by VILLUM FONDEN (grant no. 37766), and the Danish Research Foundation.
NW acknowledges support from a Royal Society - Science Foundation Ireland University Research Fellowship.
This publication has emanated from research conducted with the financial support of Science Foundation Ireland under Grant numbers 16/RS-URF/3428 and 17/RS-URF-RG/3490.
AP acknowledges support from a Royal Society University Research Fellowship.
For the purpose of Open Access, the authors have applied a CC BY public copyright licence to any Author Accepted Manuscript version arising from this submission.
This work makes use of the Black Hole Perturbation Toolkit~\cite{BHPToolkit}. Finally, the authors acknowledge the computational resources provided by the Max Planck Institute
for Gravitational Physics (Albert Einstein Institute) in Potsdam, in particular, the {\tt Hypatia} cluster.
The \texttt{SEOBNRv5} family of models is publicly available through the Python package \texttt{pySEOBNR} \href{https://git.ligo.org/lscsoft/pyseobnr}{\texttt{git.ligo.org/lscsoft/pyseobnr}}. Stable versions of \texttt{pySEOBNR} are published through the Python Package Index (PyPI), and can be installed via ~\texttt{pip install pyseobnr}.
\appendix

\section{Independence of results on the specific EOB Hamiltonian}

\begin{figure}[!tbp]
	\includegraphics[width=\columnwidth]{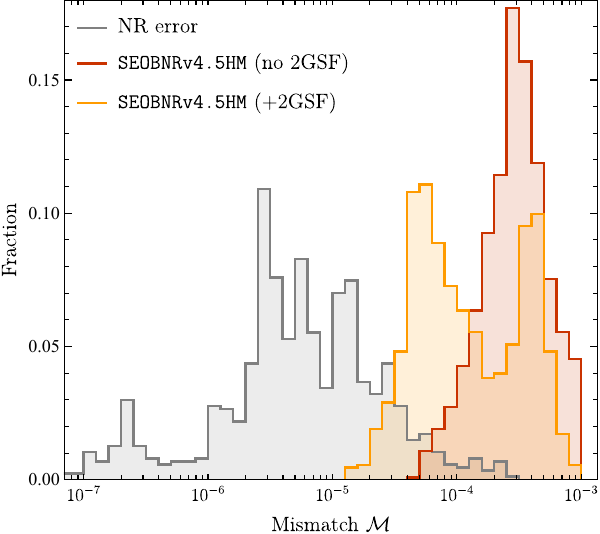}
	\caption{Similar to Fig.~\ref{fig:mismatchcal}, but now we show the mismatches for the model built by calibrating the \SEOBNR{v4} Hamiltonian,
and using the \SEOBNR{v5HM} RR force and gravitational modes with and without the 2GSF information. We label this model {\tt SEOBNRv4.5HM}. }
	\label{fig:v4mismatch}
\end{figure}
\begin{figure}[!tbp]
	\includegraphics[width=\columnwidth]{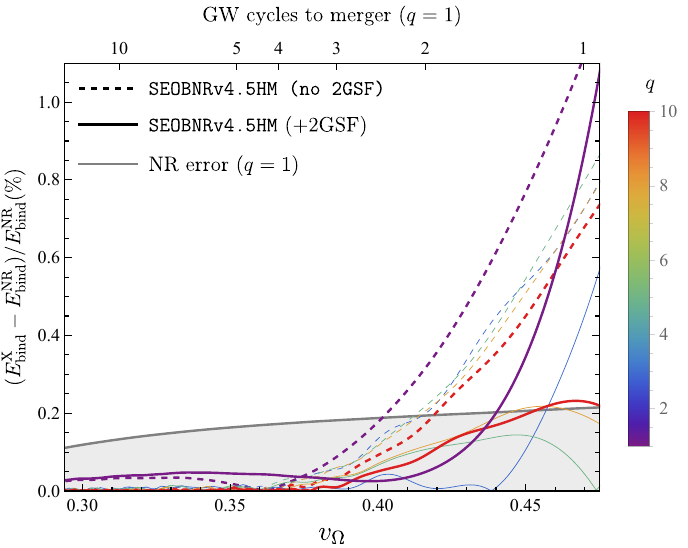}
	\caption{Similar to Fig.~\ref{fig:bindingenergy}, but now we show results for the {\tt SEOBNRv4.5HM} model with and without 2GSF corrections.
}
	\label{fig:v4bindElog}
\end{figure}

The accurate mismatches and small binding-energy disagreements with NR found in Sec.~\ref{sec:results} have been obtained
using the \texttt{SEOBNRv5} nonspinning Hamiltonian of Ref.~\cite{Khalilv5,Pompiliv5}. Here, we want to
understand whether those results are somehow tied to the particular structure (or resummation) of the
Hamiltonian and the PN content. Thus, we repeat some of the analyses using the non-spinning Hamiltonian from the previous {\tt SEOBNR} family,
\texttt{SEOBNRv4}~\cite{Taracchini:2013rva,Bohe:2016gbl}.

Figure~\ref{fig:v4mismatch} is similar to Fig.~\ref{fig:mismatchcal}, but now we compute the mismatches between the \SEOBNR{v4.5HM} and NR (2,2)
waveforms, where the \SEOBNR{v4.5HM} model is constructed calibrating the \SEOBNR{v4} Hamiltonian, and using the \SEOBNR{v5HM} RR force and gravitational modes with and without
the 2GSF information. We find that although the mismatches are a bit higher than for the \SEOBNR{v5} ones, there is actually
a noticeable improvement when including the 2GSF information.

In Fig.~\ref{fig:v4bindElog} we revisit Fig.~\ref{fig:bindingenergy} with the {\tt SEOBNRv4.5HM} model. We again see that calibrating the model using the
2GSF information in the RR force and gravitational modes leads to a more accurate recovery of the binding energy, albeit less striking than in the case
of the \SEOBNR{v5HM} model.

\section{Extended comparison between GSF and NR multipolar fluxes}\label{apdx:GSFvsNR}

\begin{figure*}[!tbp]
	\includegraphics[width=\columnwidth]{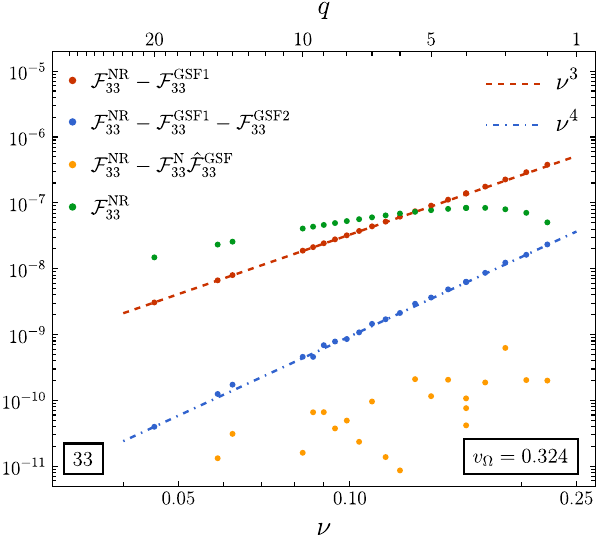}
	\includegraphics[width=\columnwidth]{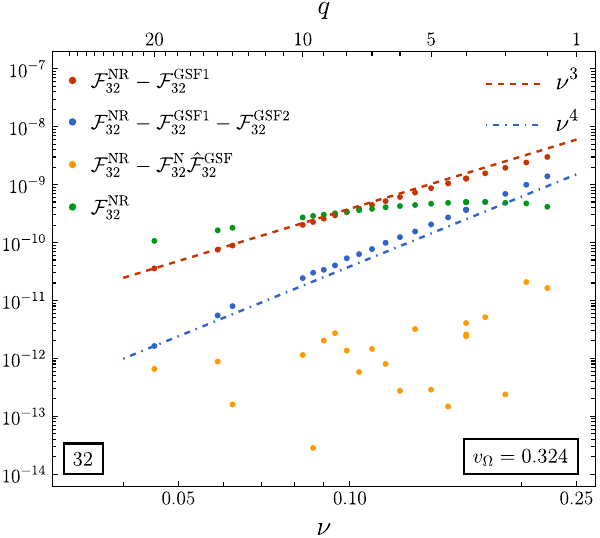}
	\includegraphics[width=\columnwidth]{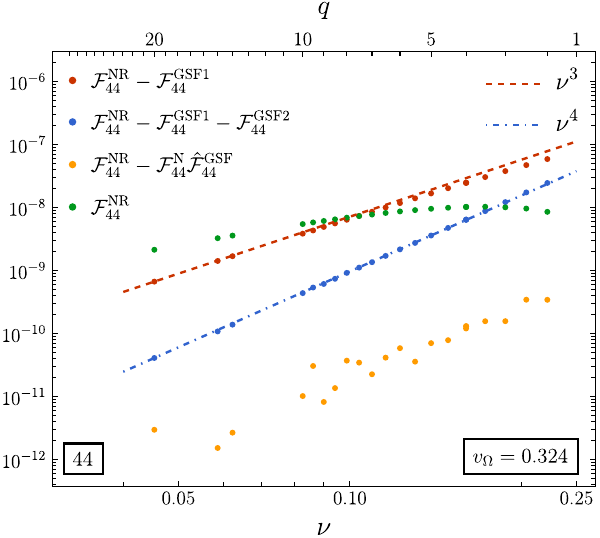}
	\includegraphics[width=\columnwidth]{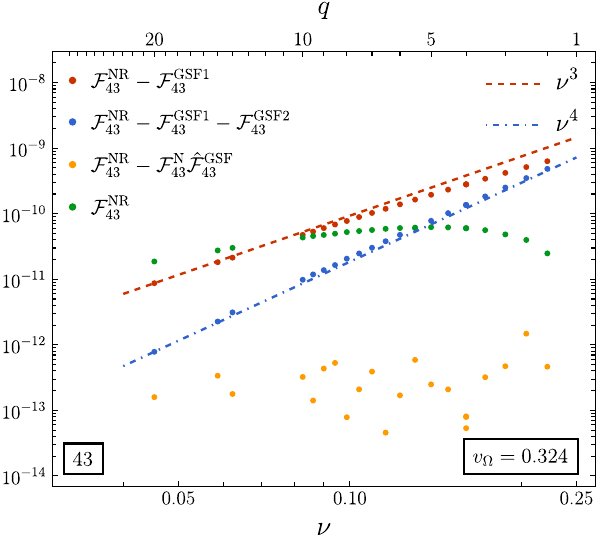}
	\caption{
		Comparison of the energy fluxes extracted from NR simulations and GSF calculations for the (3,3), (3,2), (4,4) and (4,3) modes at $r/GM \equiv 1/v_\Omega^2 = 9.5$ as a function of the symmetric mass-ratio, $\smr$. After subtracting the 1GSF (resp.~2GSF) flux from the NR flux, we see the residual scales as $\smr^3$ ($\smr^4$), as expected.}\label{fig:l3-l4-flux-scaling}
\end{figure*}

\begin{figure*}[!tbp]
	\includegraphics[width=\columnwidth]{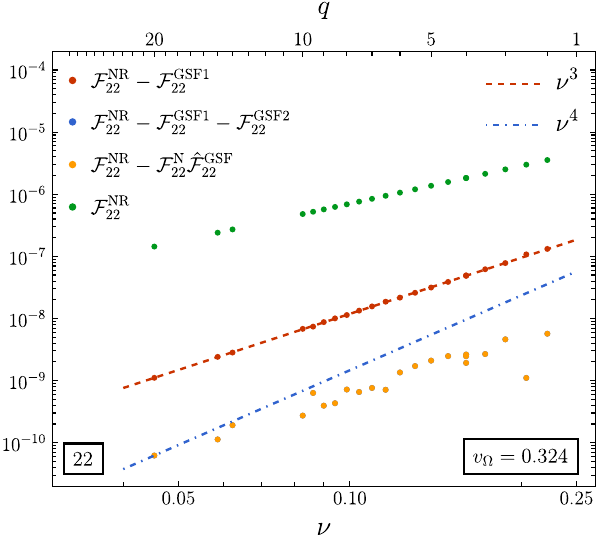}
	\includegraphics[width=\columnwidth]{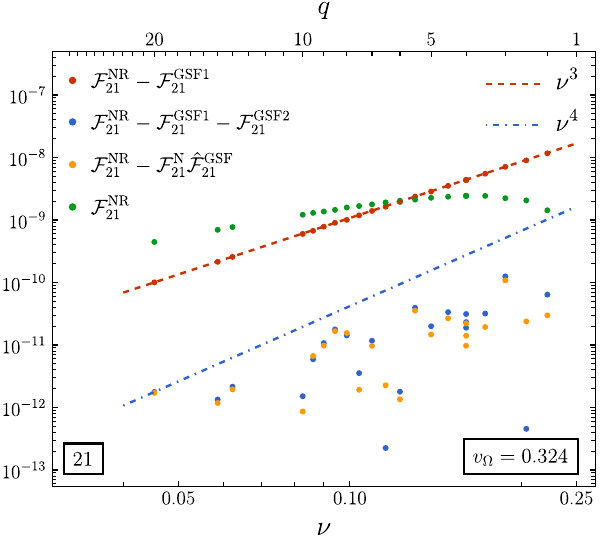}
	\caption{ Comparison of the fluxes extracted from NR
          simulations and GSF calculations for the $l=2$ modes at
          $r/GM \equiv 1/v_\Omega^2 = 9.5$ as a function of
          the symmetric mass-ratio, $\smr$. After the first-order
          flux is subtracted from the NR flux, the residual scales as
          $\smr^3$.  For the (2,1) mode after the second-order flux is
          also subtracted, the residual is within the magnitude of the
          oscillations in the NR data and so the scaling of the
          residual is less clear.  For the (2,2) mode the residual
          does not clearly follow the dash-dotted (blue) $\nu^4$ curve as
          it is likely contaminated by effects related to the
          transition to plunge.  The effect of this transition is
          clear for orbital radii close to the ISCO, as one can see in
          Fig.~\ref{fig:l2m2-flux-scaling-with-transition}.
        }\label{fig:l2-flux-scaling}
\end{figure*}

In this appendix we show further comparisons between GSF and NR multipolar energy fluxes.
The comparisons for the (3,2), (3,3), (4,3) and (4,4) modes are shown in Fig.~\ref{fig:l3-l4-flux-scaling}. The residual after subtracting the GSF flux from the NR flux clearly shows the expected scaling. The comparisons for the (2,1) and (2,2) modes are presented in Fig.~\ref{fig:l2-flux-scaling}. The scaling of the residuals for these modes is less clear for the reasons given below and in the caption of the figure. The SXS datasets used to make Figs.~\ref{fig:55-flux-scaling}, \ref{fig:l3-l4-flux-scaling}, and \ref{fig:l2-flux-scaling} are given in Table~\ref{tab:nrsims}.

\begin{figure}[!tbp]
	\includegraphics[width=\columnwidth]{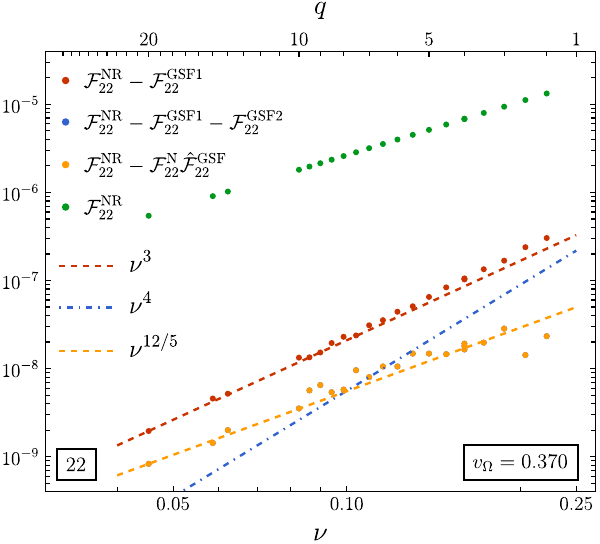}
	\caption{The same as the left panel of Fig.~\ref{fig:l2-flux-scaling}, but now for $r/GM \equiv 1/v_\Omega^2 = 7.3$.
		The dashed (yellow) curve is a $\nu^{12/5}$ reference, which is the expected scaling for the flux near the transition to plunge.
	}
	\label{fig:l2m2-flux-scaling-with-transition}
\end{figure}

Figure~\ref{fig:l2-flux-scaling} shows that for the (2,2) mode, the
flux's scaling with $\nu$ at $v_\Omega=0.324$ is likely affected by
the transition to plunge. Such a transition occurs over a frequency
interval $\sim \nu^{2/5}/M$ on a timescale $\sim M/\nu^{1/5}$, during
which the small parameter $\nu^{1/5}$ enters into the SMR expansion~\cite{Compere:2021zfj}.
The behavior of the energy flux in this case can be obtained by combining Eq.~(10) of
Ref.~\cite{Albalat:2022vcy} with Eqs.~(22) and~(23) of
Ref.~\cite{Albertini:2022rfe}. In those equations, we define
$R=(r-6M)/\nu^{2/5}$ and $\Delta\Omega=(\Omega-\Omega_{\rm
  ISCO})/\nu^{2/5}$, where $r$ is the orbital separation, such that
$R\sim M$ and $\Delta\Omega\sim 1/M$ in the transition regime. The
cited equations then give us
\begin{align}
\frac{dE}{dt} &= \frac{dE}{dR}\frac{dR}{d\Delta\Omega}\frac{d\Delta\Omega}{dt}, \nonumber\\
		&\sim (\nu^{4/5}+\nu^{6/5}+\ldots)(\nu^0+\nu^{2/5}+\ldots)\nonumber\\
		&\quad \times(\nu^{1/5}+\nu^{3/5}+\ldots),\nonumber\\
		&\sim \nu+\nu^{7/5}+\ldots .
\end{align}
Here $E$ is the specific binding energy, meaning it is related to the
flux by ${\cal F}=-\nu{dE}/{dt}$, which implies ${\cal F} \sim
\nu^2 + \nu^{12/5}+\ldots$. (In all of these schematic equations, the
reader should understand that powers of $\nu$ come with
$\Delta\Omega$-dependent coefficients, which we omit to avoid
introducing additional notation.)
Figure~\ref{fig:l2m2-flux-scaling-with-transition} repeats the
comparison in Fig.~\ref{fig:l2-flux-scaling} at a frequency closer to
the ISCO ($v_\Omega=0.370$), and there we see clear evidence of the
$\nu^{12/5}$ term appearing in the flux.

\section{Numerical-relativity simulations}\label{app:nrsims}

Throughout this paper we compare to NR simulations produced by the SXS collaboration.\footnote{See the SXS Gravitational Waveform Database \url{https://data.black-holes.org/waveforms/index.html}.}
In Table~\ref{tab:nrsims}, we provide some details of the NR simulations used in this paper.
We selected all public SXS nonspinning quasicircular simulations with sufficiently different mass ratios, and initial eccentricity below $3 \times 10^{-3}$.
If more were available for the same parameters, we took the most recent one, or the second latest if that is at least 5 orbits longer.

\begin{table*}[!tbp]
\begin{ruledtabular}
\begin{tabular}{l  l l  r r l}
SXS ID & \multicolumn{1}{c}{$\lmr$}  & \multicolumn{1}{c}{$\smr$} & \multicolumn{1}{c}{$\chi_1$} & \multicolumn{1}{c}{$\chi_2$} & Used in Fig.\\
\midrule
SXS:BBH:2325	&	1.000	&	0.2500	&	$3.64\times 10^{-5}$	&	$3.60\times 10^{-5}$	&\ref{fig:NRflux2},\ref{fig:WFplot},\ref{fig:mismatchcal},\ref{fig:v4mismatch}	\\
SXS:BBH:0198	&	1.202	&	0.2479	&	$-5.04\times 10^{-5}$	&	$8.54\times 10^{-5}$	&\ref{fig:NRflux2},\ref{fig:mismatchcal},\ref{fig:v4mismatch}\\
SXS:BBH:0310	&	1.221	&	0.2475	&	$1.46\times 10^{-4}$	&	$9.71\times 10^{-5}$	&\ref{fig:NRflux2},\ref{fig:mismatchcal},\ref{fig:v4mismatch}\\
SXS:BBH:1143	&	1.250	&	0.2469	&	$-1.37\times 10^{-4}$	&	$-2.55\times 10^{-5}$	&\ref{fig:NRflux2},\ref{fig:mismatchcal},\ref{fig:v4mismatch}\\
SXS:BBH:2331	&	1.500	&	0.2400	&	$-7.58\times 10^{-5}$	&	$-6.80\times 10^{-6}$	&\ref{fig:NRflux2},\ref{fig:mismatchcal},\ref{fig:v4mismatch}\\
SXS:BBH:0194	&	1.518	&	0.2394	&	$3.19\times 10^{-5}$	&	$-8.57\times 10^{-5}$	&\ref{fig:NRflux2},\ref{fig:mismatchcal},\ref{fig:v4mismatch}\\
SXS:BBH:1354	&	1.832	&	0.2284	&	$-1.50\times 10^{-4}$	&	$1.26\times 10^{-4}$	&\ref{fig:NRflux2},\ref{fig:mismatchcal},\ref{fig:v4mismatch}\\
SXS:BBH:1165	&	2.000 	& 	0.2222	& 	$7.91\times 10^{-5}$	& 	$1.95\times 10^{-5}$	&\ref{fig:55-flux-scaling},\ref{fig:l3-l4-flux-scaling},\ref{fig:l2-flux-scaling},\ref{fig:l2m2-flux-scaling-with-transition}  \\
SXS:BBH:2425	&	2.000	&	0.2222	&	$-7.66\times 10^{-5}$	&	$-1.16\times 10^{-4}$	&\ref{fig:NRflux2},\ref{fig:mismatchcal},\ref{fig:v4mismatch}\\
SXS:BBH:0201	&	2.316	&	0.2106	&	$6.23\times 10^{-5}$	&	$-4.16\times 10^{-5}$	&\ref{fig:NRflux2},\ref{fig:mismatchcal},\ref{fig:v4mismatch}\\
SXS:BBH:0259	&	2.500	&	0.2041	&	$9.37\times 10^{-8}$	&	$2.48\times 10^{-7}$	&\ref{fig:55-flux-scaling},\ref{fig:NRflux2},\ref{fig:mismatchcal},\ref{fig:v4mismatch},\ref{fig:l3-l4-flux-scaling},\ref{fig:l2-flux-scaling},\ref{fig:l2m2-flux-scaling-with-transition}\\
SXS:BBH:2265	&	3.000 	& 	0.1875	& 	$2.24\times 10^{-6}$ 	& 	$5.41\times 10^{-6}$ 	&\ref{fig:55-flux-scaling},\ref{fig:l3-l4-flux-scaling},\ref{fig:l2-flux-scaling},\ref{fig:l2m2-flux-scaling-with-transition}\\
SXS:BBH:2498	&	3.000	&	0.1875	&	$4.36\times 10^{-6}$	&	$3.13\times 10^{-6}$	&\ref{fig:NRflux2},\ref{fig:mismatchcal},\ref{fig:v4mismatch}\\
SXS:BBH:0200	&	3.272	&	0.1793	&	$-5.03\times 10^{-5}$	&	$-1.09\times 10^{-5}$	&\ref{fig:NRflux2},\ref{fig:mismatchcal},\ref{fig:v4mismatch}\\
SXS:BBH:2483	&	3.500	&	0.1728	&	$-2.71\times 10^{-5}$	&	$6.29\times 10^{-5}$	&\ref{fig:55-flux-scaling},\ref{fig:NRflux2},\ref{fig:mismatchcal},\ref{fig:v4mismatch},\ref{fig:l3-l4-flux-scaling},\ref{fig:l2-flux-scaling},\ref{fig:l2m2-flux-scaling-with-transition}\\
SXS:BBH:2485	&	3.999 	& 	0.1600 	& 	$2.65\times 10^{-5}$ 	& 	$8.58\times 10^{-5}$ 	& \ref{fig:55-flux-scaling},\ref{fig:l3-l4-flux-scaling},\ref{fig:l2-flux-scaling},\ref{fig:l2m2-flux-scaling-with-transition}\\
SXS:BBH:1906	&	4.000 	& 	0.1600 	& 	$5.77\times 10^{-5}$ 	& 	$8.54\times 10^{-5}$ 	& \ref{fig:55-flux-scaling},\ref{fig:l3-l4-flux-scaling},\ref{fig:l2-flux-scaling},\ref{fig:l2m2-flux-scaling-with-transition}\\
SXS:BBH:2499	&	4.000	&	0.1600	&	$8.41\times 10^{-6}$	&	$3.42\times 10^{-6}$	&\ref{fig:NRflux},\ref{fig:NRflux2},\ref{fig:WFplot},\ref{fig:mismatchcal},\ref{fig:v4mismatch}\\
SXS:BBH:1220	&	4.001 	& 	0.1600 	& 	$5.63\times 10^{-5}$ 	& 	$3.31\times 10^{-5}$ 	&\ref{fig:55-flux-scaling},\ref{fig:l3-l4-flux-scaling},\ref{fig:l2-flux-scaling},\ref{fig:l2m2-flux-scaling-with-transition}\\
SXS:BBH:2484	&	4.500	&	0.1488	&	$1.82\times 10^{-5}$	&	$-8.99\times 10^{-5}$	&\ref{fig:55-flux-scaling},\ref{fig:NRflux2},\ref{fig:mismatchcal},\ref{fig:v4mismatch},\ref{fig:l3-l4-flux-scaling},\ref{fig:l2-flux-scaling},\ref{fig:l2m2-flux-scaling-with-transition}\\
SXS:BBH:2374	&	5.000	&	0.1389	&	$-8.13\times 10^{-5}$	&	$5.24\times 10^{-5}$	&\ref{fig:NRflux2},\ref{fig:mismatchcal},\ref{fig:v4mismatch}\\
SXS:BBH:2487	&	5.000 	& 	0.1389 	& 	$8.38\times 10^{-6}$ 	& 	$1.53\times 10^{-5}$ 	&\ref{fig:55-flux-scaling},\ref{fig:l3-l4-flux-scaling},\ref{fig:l2-flux-scaling},\ref{fig:l2m2-flux-scaling-with-transition}\\
SXS:BBH:0187	&	5.039	&	0.1381	&	$8.80\times 10^{-6}$	&	$-1.20\times 10^{-5}$	&\ref{fig:NRflux2},\ref{fig:mismatchcal},\ref{fig:v4mismatch}\\
SXS:BBH:2486	&	5.500	&	0.1302	&	$-2.80\times 10^{-6}$	&	$-9.81\times 10^{-6}$	&\ref{fig:55-flux-scaling},\ref{fig:NRflux2},\ref{fig:mismatchcal},\ref{fig:v4mismatch},\ref{fig:l3-l4-flux-scaling},\ref{fig:l2-flux-scaling},\ref{fig:l2m2-flux-scaling-with-transition}\\
SXS:BBH:0197	&	5.522	&	0.1298	&	$-3.70\times 10^{-5}$	&	$-1.52\times 10^{-5}$	&\ref{fig:NRflux2},\ref{fig:mismatchcal},\ref{fig:v4mismatch}\\
SXS:BBH:2489	&	5.999 	& 	0.1225 	& 	$8.03\times 10^{-6}$ 	& 	$3.72\times 10^{-5}$ 	&\ref{fig:55-flux-scaling},\ref{fig:l3-l4-flux-scaling},\ref{fig:l2-flux-scaling},\ref{fig:l2m2-flux-scaling-with-transition}\\
SXS:BBH:2164	&	6.000	&	0.1225	&	$-2.71\times 10^{-6}$	&	$-1.42\times 10^{-5}$	&\ref{fig:NRflux2},\ref{fig:mismatchcal},\ref{fig:v4mismatch}\\
SXS:BBH:2488	&	6.500	&	0.1155	&	$2.79\times 10^{-5}$	&	$-2.41\times 10^{-5}$	&\ref{fig:55-flux-scaling},\ref{fig:NRflux2},\ref{fig:mismatchcal},\ref{fig:v4mismatch},\ref{fig:l3-l4-flux-scaling},\ref{fig:l2-flux-scaling},\ref{fig:l2m2-flux-scaling-with-transition}\\
SXS:BBH:0192	&	6.580	&	0.1145	&	$2.51\times 10^{-5}$	&	$-5.07\times 10^{-5}$	&\ref{fig:NRflux2},\ref{fig:mismatchcal},\ref{fig:v4mismatch}\\
SXS:BBH:2491	&	7.000	&	0.1094	&	$1.14\times 10^{-5}$	&	$4.51\times 10^{-5}$	&\ref{fig:55-flux-scaling},\ref{fig:NRflux2},\ref{fig:mismatchcal},\ref{fig:v4mismatch},\ref{fig:l3-l4-flux-scaling},\ref{fig:l2-flux-scaling},\ref{fig:l2m2-flux-scaling-with-transition}\\
SXS:BBH:0188	&	7.187	&	0.1072	&	$1.55\times 10^{-6}$	&	$-2.45\times 10^{-5}$	&\ref{fig:NRflux2},\ref{fig:mismatchcal},\ref{fig:v4mismatch}\\
SXS:BBH:2490	&	7.500	&	0.1038	&	$-2.92\times 10^{-5}$	&	$-5.94\times 10^{-6}$	&\ref{fig:55-flux-scaling},\ref{fig:mismatchcal},\ref{fig:v4mismatch},\ref{fig:l3-l4-flux-scaling},\ref{fig:l2-flux-scaling},\ref{fig:l2m2-flux-scaling-with-transition}\\
SXS:BBH:0195	&	7.761	&	0.1011	&	$1.32\times 10^{-5}$	&	$-4.01\times 10^{-5}$	&\ref{fig:NRflux2},\ref{fig:mismatchcal},\ref{fig:v4mismatch}\\
SXS:BBH:2493	&	8.000	&	0.09876	&	$2.68\times 10^{-5}$	&	$-4.49\times 10^{-5}$	&\ref{fig:55-flux-scaling},\ref{fig:NRflux2},\ref{fig:mismatchcal},\ref{fig:v4mismatch},\ref{fig:l3-l4-flux-scaling},\ref{fig:l2-flux-scaling},\ref{fig:l2m2-flux-scaling-with-transition}\\
SXS:BBH:0186	&	8.267	&	0.09626	&	$1.02\times 10^{-6}$	&	$-8.82\times 10^{-8}$	&\ref{fig:NRflux2},\ref{fig:mismatchcal},\ref{fig:v4mismatch}\\
SXS:BBH:2492	&	8.501	&	0.09417	&	$-3.20\times 10^{-6}$	&	$-1.83\times 10^{-5}$	&\ref{fig:55-flux-scaling},\ref{fig:NRflux2},\ref{fig:mismatchcal},\ref{fig:v4mismatch},\ref{fig:l3-l4-flux-scaling},\ref{fig:l2-flux-scaling},\ref{fig:l2m2-flux-scaling-with-transition}\\
SXS:BBH:0199	&	8.729	&	0.09222	&	$-1.11\times 10^{-6}$	&	$-3.31\times 10^{-5}$	&\ref{fig:NRflux2},\ref{fig:mismatchcal},\ref{fig:v4mismatch}\\
SXS:BBH:2495	&	9.001	&	0.08999	&	$1.36\times 10^{-6}$	&	$-8.77\times 10^{-6}$	&\ref{fig:55-flux-scaling},\ref{fig:NRflux2},\ref{fig:mismatchcal},\ref{fig:v4mismatch},\ref{fig:l3-l4-flux-scaling},\ref{fig:l2-flux-scaling},\ref{fig:l2m2-flux-scaling-with-transition}\\
SXS:BBH:0189	&	9.167	&	0.08868	&	$1.18\times 10^{-5}$	&	$-6.79\times 10^{-6}$	&\ref{fig:NRflux2},\ref{fig:mismatchcal},\ref{fig:v4mismatch}\\
SXS:BBH:1108	&	9.200	&	0.08843	&	$-2.25\times 10^{-6}$	&	$-1.46\times 10^{-6}$	&\ref{fig:NRflux2},\ref{fig:mismatchcal},\ref{fig:v4mismatch}\\
SXS:BBH:2494	&	9.497	&	0.08619	&	$-1.57\times 10^{-5}$	&	$-3.54\times 10^{-5}$	&\ref{fig:55-flux-scaling},\ref{fig:NRflux2},\ref{fig:mismatchcal},\ref{fig:v4mismatch},\ref{fig:l3-l4-flux-scaling},\ref{fig:l2-flux-scaling},\ref{fig:l2m2-flux-scaling-with-transition}\\
SXS:BBH:0196	&	9.663	&	0.08499	&	$1.67\times 10^{-6}$	&	$-2.73\times 10^{-5}$	&\ref{fig:NRflux2},\ref{fig:mismatchcal},\ref{fig:v4mismatch}\\
SXS:BBH:0185	&	9.990	&	0.08271	&	$1.28\times 10^{-5}$	&	$-1.43\times 10^{-5}$	&\ref{fig:NRflux2},\ref{fig:WFplot},\ref{fig:mismatchcal},\ref{fig:v4mismatch}\\
SXS:BBH:1107	&	10.00 	& 	0.08264	& 	$3.66\times 10^{-6}$	& 	$1.06\times 10^{-7}$	&\ref{fig:55-flux-scaling},\ref{fig:l3-l4-flux-scaling},\ref{fig:l2-flux-scaling},\ref{fig:l2m2-flux-scaling-with-transition}\\
SXS:BBH:2480	&	14.00	&	0.06222	&	$7.62\times 10^{-6}$	&	$-4.14\times 10^{-6}$	&\ref{fig:55-flux-scaling},\ref{fig:NRflux2},\ref{fig:mismatchcal},\ref{fig:v4mismatch},\ref{fig:l3-l4-flux-scaling},\ref{fig:l2-flux-scaling},\ref{fig:l2m2-flux-scaling-with-transition}\\
SXS:BBH:2477	&	15.00	&	0.05859	&	$6.43\times 10^{-6}$	&	$-4.52\times 10^{-6}$	&\ref{fig:55-flux-scaling},\ref{fig:NRflux2},\ref{fig:mismatchcal},\ref{fig:v4mismatch},\ref{fig:l3-l4-flux-scaling},\ref{fig:l2-flux-scaling},\ref{fig:l2m2-flux-scaling-with-transition}\\
SXS:BBH:2516	&	20.00	&	0.04536	&	$3.43\times 10^{-5}$	&	$-1.02\times 10^{-4}$	&\ref{fig:55-flux-scaling},\ref{fig:NRflux2},\ref{fig:mismatchcal},\ref{fig:v4mismatch},\ref{fig:l3-l4-flux-scaling},\ref{fig:l2-flux-scaling},\ref{fig:l2m2-flux-scaling-with-transition}
\end{tabular}
\end{ruledtabular}

\caption{Details of the SXS simulations used in Figures throughout the paper.}
\label{tab:nrsims}
\end{table*}

\bibliography{references}

\end{document}